\documentclass[conference, twocolumn]{IEEEtran}

\IEEEoverridecommandlockouts
\usepackage{amsmath,amssymb,amsfonts}
\usepackage{graphicx}
\usepackage{textcomp}
\usepackage{xcolor}
\usepackage{bm}
\usepackage[maxcitenames=6,
  doi=false,
  url=false,
  isbn=false,
  style=ieee,
  citestyle=ieee-comp,
  sortcites=true]{biblatex}
\usepackage{tikz}
\usepackage{comment}
\usepackage{tikz-3dplot}
\usepackage{keyval}
\usepackage{ifthen}
\usepackage{acro}
\usepackage{algorithm}
\usepackage{algpseudocode}
\usepackage{siunitx}
\usepackage{booktabs}
\ifCLASSOPTIONcompsoc
\usepackage[caption=false,font=normalsize,labelfon
t=sf,textfont=sf]{subfig}
\else
\usepackage[caption=false,font=footnotesize]{subfig}

\fi
\newif\ifarxiv
\arxivtrue
\newcommand{\arxivonly}[1]{%
    \ifarxiv
        #1%
    \fi
}
\newcommand{\versiontext}[2]{%
    \ifarxiv
        #1%
    \else
        #2%
    \fi
}
\makeatletter
\newcommand{\figwidth}{%
  \if@twocolumn
    0.8\linewidth
  \else
    5in
  \fi
}
\newcommand{\linebreaker}[2]{%
  \if@twocolumn
    #1%
  \else
    #2%
  \fi
}
\makeatother
\DeclareAcronym{3d}{
        short = {3D},
        long = {three-dimensional},
}
\DeclareAcronym{isac}{
        short = {ISAC},
        long = {integrated sensing and communication},
}
\DeclareAcronym{nf}{
        short = {NF},
        long = {near-field},
}
\DeclareAcronym{bs}{
        short = {BS},
        long = {base station},
}
\DeclareAcronym{ma}{
        short = {MA},
        long = {movable antenna},
        long-plural-form = {movable antennas}
}
\DeclareAcronym{roi}{
        short = {ROI},
        long = {region of interest},
}
\DeclareAcronym{los}{
        short = {LoS},
        long = {line-of-sight},
}
\DeclareAcronym{wrt}{
        short = {w.r.t.},
        long = {with respect to},
}
\DeclareAcronym{2d}{
        short = {2D},
        long = {two-dimensional},
}
\DeclareAcronym{awgn}{
        short = {AWGN},
        long = {additive white Gaussian noise},
}
\DeclareAcronym{psf}{
        short = {PSF},
        long = {point spread function},
}

\DeclareAcronym{isl}{
        short = {ISL},
        long = {integrated sidelobe level},
}
\DeclareAcronym{psl}{
        short = {PSL},
        long = {peak sidelobe level},
}
\DeclareAcronym{ao}{
        short = {AO},
        long = {alternating optimization},
}
\DeclareAcronym{ga}{
        short = {GA},
        long = {genetic algorithm},
}
\DeclareAcronym{pso}{
        short = {PSO},
        long = {particle swarm optimization},
}
\DeclareAcronym{mmwave}{
        short = {mmWave},
        long = {millimeter-wave},
}
\DeclareAcronym{mse}{
        short = {MSE},
        long = {mean-squared error},
}
\DeclareAcronym{mimo}{
        short = {MIMO},
        long = {multiple-input multiple-output},
}
\DeclareAcronym{dof}{
        short = {DoF},
        long = {degrees of freedom},
}
\DeclareAcronym{sinr}{
        short = {SINR},
        long = {signal-to-interference-plus-noise ratio},
}
\DeclareAcronym{zf}{
        short = {ZF},
        long = {zero-forcing},
}
\DeclareAcronym{v2x}{
        short = {V2X},
        long = {vehicle-to-everything},
}
\DeclareAcronym{crb}{
        short = {CRB},
        long = {Cram\'er-Rao bound},
}
\DeclareAcronym{cr}{
        short = {CR},
        long = {communication-rate},
}
\DeclareAcronym{ff}{
        short = {FF},
        long = {far-field},
}
\DeclareAcronym{6g}{
        short = {6G},
        long = {sixth-generation},
}
\DeclareAcronym{rcs}{
        short = {RCS},
        long = {radar cross section},
}

\usetikzlibrary{shapes,arrows,positioning,calc,automata,positioning,fit,shapes.geometric}
\usetikzlibrary{arrows.meta,calc,positioning,decorations.pathreplacing,shapes.geometric}
\usetikzlibrary{angles,quotes} 
\def\BibTeX{{\rm B\kern-.05em{\sc i\kern-.025em b}\kern-.08em
    T\kern-.1667em\lower.7ex\hbox{E}\kern-.125emX}}
\AtEveryBibitem{\clearlist{language}}

\begin{document}
\title{Joint Antenna Position and Beamforming Optimization for Movable Antenna Enabled Near-Field ISAC Imaging
}

\linebreaker{\author{\IEEEauthorblockN{Luis Ebner\IEEEauthorrefmark{1}, Alireza Navi\IEEEauthorrefmark{1}, Mohamed-Amine~Lahmeri\IEEEauthorrefmark{1}, \\ Sebastian Lotter\IEEEauthorrefmark{1}, Lutz Lampe\IEEEauthorrefmark{2},  and Robert Schober\IEEEauthorrefmark{1}}\\[-1.0em]
\IEEEauthorblockA{\IEEEauthorrefmark{1}Friedrich-Alexander-Universit\"at Erlangen-N\"urnberg, Germany
}\\[-1.0em]
\IEEEauthorblockA{\IEEEauthorrefmark{2}University of British Columbia, Canada}}
}{\author{\IEEEauthorblockN{Luis Ebner\IEEEauthorrefmark{1}, Alireza Navi\IEEEauthorrefmark{1}, Mohamed-Amine~Lahmeri\IEEEauthorrefmark{1}, \\ Sebastian Lotter\IEEEauthorrefmark{1}, Lutz Lampe\IEEEauthorrefmark{2},  and Robert Schober\IEEEauthorrefmark{1}}\\[-1.8em]
\IEEEauthorblockA{\IEEEauthorrefmark{1}Friedrich-Alexander-Universit\"at Erlangen-N\"urnberg, Germany
}\\[-1.8em]
\IEEEauthorblockA{\IEEEauthorrefmark{2}University of British Columbia, Canada}}
}
\maketitle
\begin{abstract}
Near-field (NF) integrated sensing and communication (ISAC) systems enable simultaneous two-dimensional (2D) radar imaging and multi-user downlink communication. However, fixed-position antenna arrays can exhibit high sidelobes in the point spread function (PSF), which degrade imaging performance and cause ghost-target artifacts. Movable antennas (MAs) offer additional spatial degrees of freedom to suppress these sidelobes while meeting communication requirements, without increasing the number of antenna elements. In this paper, we investigate MA-enabled NF ISAC by jointly optimizing the transmit and receive antenna positions and the transmit beamforming to minimize a weighted sum of the integrated sidelobe level (ISL) and peak sidelobe level (PSL) of the 2D PSF, subject to communication-rate, illumination-power, transmit-power, and antenna-placement constraints. To tackle the resulting non-convex problem, we propose a hybrid alternating optimization (AO) framework that combines particle swarm optimization with a genetic algorithm. Our numerical results demonstrate that the proposed scheme achieves the lowest ISL among the considered array configurations and reduces the PSL by at least 5 dB compared with the benchmarks. The optimized array is further shown to resolve two closely spaced targets whose reflectivities differ by 6dB, outperforming random, Mills Cross, and rectangular arrays while satisfying the imposed minimum user-throughput requirements.
\end{abstract}


\section{Introduction}
\Ac{isac} has emerged as a key technology for future \ac{6g} networks by sharing hardware, spectrum, and signal-processing resources. It enables both functionalities for applications such as vehicle-to-everything, smart manufacturing, and environmental monitoring \cite{liuIntegratedSensingCommunications2022}. Conventionally, \ac{mimo} \ac{isac} approaches primarily focus on parameter-estimation-oriented sensing, targeting quantities such as range, velocity, and angle. Commonly adopted sensing objectives include beampattern matching, the \ac{crb}, and radar signal-quality metrics \cite{liuIntegratedSensingCommunications2022, quNearFieldIntegratedSensing2024, rahalNearFieldWidebandISAC2025}. Unlike parameter estimation, radar imaging aims to reconstruct the spatial reflectivity distribution of an extended scene. Its performance therefore depends on the spatial response of the imaging system, namely its resolution and sidelobe characteristics. Sidelobes and grating lobes can introduce imaging artifacts and interfere with the reconstruction of neighboring scatterers \cite{zhugeStudyTwoDimensionalSparse2012, ahmedElectronicMicrowaveImaging}. Conventional parameter-estimation-oriented sensing metrics do not directly characterize these imaging effects. Instead, the \ac{psf} describes the reconstructed response to a point scatterer and characterizes resolution, sidelobe behavior, and spatially dependent imaging artifacts \cite{ahmedElectronicMicrowaveImaging}. In addition, high-resolution sensing and imaging typically require large apertures and high carrier frequencies, which can place the \ac{roi} within the \ac{nf} of the radar system \cite{quNearFieldIntegratedSensing2024, ahmedElectronicMicrowaveImaging}. Since antenna geometry and transmit beamforming both influence the \ac{psf} and the communication performance, their joint optimization is particularly relevant for \ac{nf} \ac{isac} systems. 

The \ac{nf} radar imaging literature has investigated sparse
antenna-array optimization to improve spatial resolution and
suppress imaging artifacts \cite{zhugeStudyTwoDimensionalSparse2012, jiangMASCOTMaskAssistedSpatial2025}. However, these approaches do not consider the joint design of antenna geometry and transmit beamforming under communication requirements. In the context of integrated imaging and communication, the authors of \cite{abusanadJointImagingDownlink2025} formulated a shared-resource optimization problem that designs the transmit beamformer to minimize the imaging and communication \ac{mse}. Nevertheless, the antenna geometry remained fixed. On the other hand, \acp{ma} introduce additional spatial \ac{dof} by allowing antenna positions to be optimized within a prescribed deployment region. Recent studies have jointly optimized \ac{ma} positions and beamforming for \ac{ff} and \ac{nf} \ac{isac} systems \cite{lyuMovableAntennaEnabled2025,dingMovableAntennaAidedNearField2026}. However, the considered parameter-estimation-based objectives do not address the sidelobes and imaging artifacts in reconstructed scenes.

To fill this gap, this paper treats the \ac{2d} \ac{psf} itself as the design objective. We use the antenna mobility to reshape the imaging response of the array directly, while guaranteeing multi-user communication performance requirements. The main contributions are as follows:
\begin{itemize}
\item To the best of the authors’ knowledge, this is the first work to optimize the \ac{ma} positions for \ac{psf}-based \ac{nf} radar imaging. To this end, we derive a \ac{psf} matrix that explicitly couples antenna geometry and transmit beamforming.
\item We formulate the joint antenna-position and beamforming optimization problem to minimize a weighted sum of the \ac{isl} and the \ac{psl} under communication, illumination, power, and antenna-placement constraints. 
\item To solve this problem, we propose an \ac{ao} framework combining \ac{pso} and \ac{ga} with feasibility-aware updates.
\item Through numerical simulations, we investigate the convergence of the proposed algorithm and
compare its imaging performance with several fixed-array benchmarks.
\end{itemize}

{\em Notations}: 
In this paper, $\bm{\mathrm{I}}_{N}$ is the identity matrix of size $N \times N$. $\{a, ..., b\}$ denotes the set of all integers between $a$ and $b$. For real numbers $a$ and $b$, $[a,b]$ and $[a,b]^{N}$ denote the closed interval $\{x \in \mathbb{R} \mid a \leq x \leq b\}$ and the set of all $N$-dimensional vectors with entries in this interval, respectively. $\mathbb{R}^{N}$ and $\mathbb{C}^{N}$ represent the sets of all $N$-dimensional vectors with real- and complex-valued entries, respectively, while $\mathbb{R}^{N \times M}$ and $\mathbb{C}^{N \times M}$ represent the sets of all $N \times M$ matrices with real- and complex-valued entries, respectively. \([\cdot]_m\) and $\log_n(\cdot)$ are the \(m\)-th element of a vector and the logarithm to base $n$, respectively. \arxivonly{$[x]_+$ refers to $\max\{0,x\}$.} Furthermore, $|\cdot|$, $||\cdot||$, and $||\cdot||_{\mathrm{F}}$ stand for the absolute value, the Euclidean norm, and the Frobenius norm, respectively. $(\cdot)^\top$, $(\cdot)^*$, and $(\cdot)^H$ represent the transpose, complex-conjugate, and Hermitian, respectively. 
\versiontext{Finally, $\operatorname{vec}(\cdot)$,  $\operatorname{diag}(\cdot)$, $\mathrm{tr}(\cdot)$, $\bullet$, $\otimes$ and $ \odot $ denote vectorization, diagonal matrix, trace, element-wise multiplication, Kronecker product, and Khatri-Rao product, respectively.}{Finally, $\operatorname{vec}(\cdot)$,  $\operatorname{diag}(\cdot)$, $\mathrm{tr}(\cdot)$, $\otimes$ and $ \odot $ denote vectorization, diagonal matrix, trace, Kronecker product, and Khatri-Rao product, respectively.}
\section{System Model}
\vspace{-0.1cm}
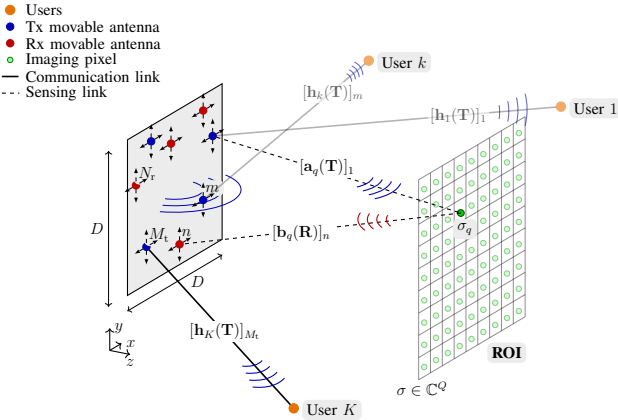
\begin{figure}[h]
\centering
\tdplotsetmaincoords{70}{30}
\resizebox{\linebreaker{0.46\textwidth}{5in}}{!}{
    \begin{tikzpicture}[tdplot_main_coords,scale=0.62]

    \tikzset{
        panel/.style={fill=gray!15},
        tx/.style={circle, draw=blue!70!black, fill=blue!70!black, minimum size=5.5pt, inner sep=0pt},
        rx/.style={circle, draw=red!75!black, fill=red!75!black, minimum size=5.5pt, inner sep=0pt},
        user/.style={circle, draw=orange!90!black, fill=orange!90!black, minimum size=7pt, inner sep=0pt},
        focus/.style={circle, draw=green!50!black, fill=green!40, minimum size=4pt, inner sep=0pt},
        voxelhi/.style={circle, draw=green!30!black, fill=green!80!black, minimum size=5pt, inner sep=0pt},
        dim/.style={<->,thin},
        comm/.style={very thick},
        radar/.style={thick,dashed},
        si/.style={thick,dotted},
        note/.style={rounded corners, font=\normalsize, fill=gray!15},
        eqlabel/.style={font=\small, fill=white, inner sep=1.5pt}
    }

    \pgfmathsetmacro{\cubex}{9}
    \pgfmathsetmacro{\cubey}{9}
    \pgfmathsetmacro{\cubez}{9}
    \def\Dx{7}         
    \def\Dy{8}         
    \def\za{13.5}      

    \coordinate (A) at (0,-\Dy/2,-\Dx/2);
    \coordinate (B) at (0, \Dy/2,-\Dx/2);
    \coordinate (C) at (0, \Dy/2, \Dx/2);
    \coordinate (D) at (0,-\Dy/2, \Dx/2);

    \coordinate (BS) at (0,0,0);
    \coordinate (O) at (0,-5.5,-5.5);
    
    \coordinate (S) at (\za,0,0);

    \coordinate (V) at (\za,0.8,1.15);



    \foreach \x in {0,...,9}
        {
            \draw[opacity=0.45]  (\za+1,\x -\cubey/2,-\cubex/2) -- (\za+1,\x -\cubey/2,\cubex/2);
        }

    \foreach \x in {0,...,9}
        {
            \draw[opacity=0.45]  (\za+1,-\cubey/2,\x - \cubex/2 ) -- (\za+1,\cubey/2,\x - \cubex/2);
        }

    \foreach \x in {0,...,8}
        \foreach \y in {0,...,8}
            \node[focus, opacity=0.45] at (\za,\x - \cubey/2 + 2.2,\y - \cubex/2 - 0.3) {};

    \node[voxelhi] at (V) {};
    \node[note, anchor=west, font=\large] at ($(V)+(0,-0.7,-0.5)$) {$\sigma_q$};

    \node[note, font=\large] at ($(S)+(+5.3,-\cubey/2,-\cubex/2+2)$) {\textbf{ROI}};
    \node[eqlabel, anchor=west, font=\large] at ($(S)+(0,-4.8,-5)$) {$\mathbf{\sigma}\in\mathbb{C}^{Q}$};

    \fill[panel,opacity=0.95] (A)--(B)--(C)--(D)--cycle;
    \draw[thick] (A)--(B)--(C)--(D)--cycle;
    \tdplotdrawarc[thin, color=blue!70!black]{(0,0,0)}{2.8}{-73}{70}{}{};
    \tdplotdrawarc[thin, color=blue!70!black]{(0,0,0)}{2}{-75}{75}{}{};
    \tdplotdrawarc[thin, color=blue!70!black]{(0,0,0)}{1.2}{-80}{80}{}{};
    \draw[->,thick] (O) -- ++(1,0,0) node[below, font=\large] {$z$};
    \draw[->,thick] (O) -- ++(0,1,0) node[right, font=\large] {$x$};
    \draw[->,thick] (O) -- ++(0,0,1) node[right, font=\large] {$y$};

    \draw[dim] ($(A)!-0.1!(D)$) -- node[below right=-3mm and 3mm, font=\large] {$D$} ($(B)!-0.1!(C)$);
    \draw[dim] ($(A)!-0.2!(B)$) -- node[left=0pt, font=\large] {$D$} ($(D)!-0.2!(C)$);

    \node[tx] (TX1) at (0,  0.8*\Dy/2,  0.40*\Dx/2) {};
    \node[tx] (TX2) at (0,  0.6*\Dy/2, -0.35*\Dx/2) {};
    \node[note, anchor=west, font=\large] at ($(TX2)+(0,-0.2,0.6)$) {$m$};
    \node[tx] (TX3) at (0, -0.5*\Dy/2,  0.8*\Dx/2) {};
    
    \node[tx] (TX4) at (0, -0.6*\Dy/2, -0.52*\Dx/2) {};
    \node[note, anchor=west, font=\large] at ($(TX4)+(0,-0.2,0.6)$) {$M_{\mathrm{t}}$};
    \foreach \tx in {TX1, TX2, TX3, TX4}
    {
        \draw[-latex,dashed] (\tx) -- ++(0,-1,0) {};
        \draw[-latex,dashed] (\tx) -- ++(0,1,0) {};
        \draw[-latex,dashed] (\tx) -- ++(0,0,0.8) {};
        \draw[-latex,dashed] (\tx) -- ++(0,0,-0.8) {};
    }

    \node[rx] (RX1) at (0,  0.6*\Dy/2,  0.8*\Dx/2) {};
    \node[rx] (RX2) at (0, -0.08*\Dy/2, 0.62*\Dx/2) {};
    \node[rx] (RX3) at (0,  0.1*\Dy/2, -0.74*\Dx/2) {};
    \node[note, anchor=west, font=\large] at ($(RX3)+(0,-0.2,0.6)$) {$n$};
    \node[rx] (RX4) at (0, -0.82*\Dy/2,  0.35*\Dx/2) {};
    \node[note, anchor=west, font=\large] at ($(RX4)+(0,-0.2,0.6)$) {$N_{\mathrm{r}}$};
    \foreach \rx in {RX1, RX2, RX3, RX4}
    {
        \draw[-latex,dashed] (\rx) -- ++(0,-1,0) {};
        \draw[-latex,dashed] (\rx) -- ++(0,1,0) {};
        \draw[-latex,dashed] (\rx) -- ++(0,0,0.8) {};
        \draw[-latex,dashed] (\rx) -- ++(0,0,-0.8) {};
    }


    \coordinate (U1c) at (20,-2,8);
    \coordinate (U2c) at (6,6,5);
    \coordinate (U3c) at (11,-9,-5);

    \node[user,opacity=0.6, font=\large] (U1) at (U1c) {};
    \node[user,opacity=0.6, font=\large] (U2) at (U2c) {};
    \node[user, font=\large] (U3) at (U3c) {};

    \node[note, anchor=west, font=\large] at ($(U1)+(0.4,0,0)$) {User $1$};
    \node[note, anchor=west, font=\large] at ($(U2)+(0.4,0,0)$) {User $k$};
    \node[note, anchor=west, font=\large] at ($(U3)+(0.4,0,0)$) {User $K$};

    \draw[comm, opacity=0.3] (TX1) -- (U1);
    \draw[comm, opacity=0.3] (TX2) -- (U2);
    \draw[comm] (TX4) -- (U3);

    \node[eqlabel,opacity=0.6, font=\large] (H1) at ($(TX1)!0.7!(U1)+(0.1,0,0.0)$) {$[\bm{\mathrm{h}}_1(\bm{\mathrm{T}})]_{1}$}; 
    \tdplotsetrotatedcoords{-45}{90}{90};
    \tdplotdrawarc[thick, color=blue!70!black,tdplot_rotated_coords, opacity=0.6]{(H1)}{1.8}{-10}{10}{}{};
    \tdplotdrawarc[thick, color=blue!70!black,tdplot_rotated_coords, opacity=0.6]{(H1)}{2.2}{-12}{12}{}{};
    \tdplotdrawarc[thick, color=blue!70!black,tdplot_rotated_coords, opacity=0.6]{(H1)}{2.6}{-13}{13}{}{};
    \tdplotdrawarc[thick, color=blue!70!black,tdplot_rotated_coords, opacity=0.6]{(H1)}{3}{-14}{14}{}{};
    
    \node[eqlabel,opacity=0.6, font=\large] (H2) at ($(TX2)!0.772!(U2)+(0.1,0,0.0)$) {$[\bm{\mathrm{h}}_k(\bm{\mathrm{T}})]_{m}$};
    \tdplotsetrotatedcoords{60}{-45}{-15};
    \tdplotdrawarc[thick, color=blue!70!black,tdplot_rotated_coords, opacity=0.6]{(H2)}{1.6}{-20}{20}{}{};
    \tdplotdrawarc[thick, color=blue!70!black,tdplot_rotated_coords, opacity=0.6]{(H2)}{1.4}{-20}{20}{}{};
    \tdplotdrawarc[thick, color=blue!70!black,tdplot_rotated_coords, opacity=0.6]{(H2)}{1.2}{-20}{20}{}{};
    \tdplotdrawarc[thick, color=blue!70!black,tdplot_rotated_coords, opacity=0.6]{(H2)}{1}{-20}{20}{}{};
    
    \node[eqlabel, font=\large] (H3) at ($(TX4)!0.52!(U3)+(0.1,0,0.0)$) {$[\bm{\mathrm{h}}_K(\bm{\mathrm{T}})]_{M_{\mathrm{t}}}$}; 
    \tdplotsetrotatedcoords{0}{35}{-10};
    \tdplotdrawarc[thick, color=blue!70!black,tdplot_rotated_coords]{(H3)}{2}{-20}{20}{}{};
    \tdplotdrawarc[thick, color=blue!70!black,tdplot_rotated_coords]{(H3)}{3.2}{-20}{20}{}{};
    \tdplotdrawarc[thick, color=blue!70!black,tdplot_rotated_coords]{(H3)}{2.8}{-20}{20}{}{};
    \tdplotdrawarc[thick, color=blue!70!black,tdplot_rotated_coords]{(H3)}{2.4}{-20}{20}{}{};

    \draw[radar] (TX1) -- (V);
    \draw[radar] (V) -- (RX3);

    \node[eqlabel, font=\large] (A1) at ($(TX1)!0.45!(V)+(-0.1,0.4,0.0)$) {$[\bm{\mathrm{a}}_q(\bm{\mathrm{T}})]_{1}$}; 
    \tdplotsetrotatedcoords{0}{10}{0};
    \tdplotdrawarc[thick, color=blue!70!black,tdplot_rotated_coords]{(A1)}{2}{-20}{20}{}{};
    \tdplotdrawarc[thick, color=blue!70!black,tdplot_rotated_coords]{(A1)}{3.2}{-20}{20}{}{};
    \tdplotdrawarc[thick, color=blue!70!black,tdplot_rotated_coords]{(A1)}{2.8}{-20}{20}{}{};
    \tdplotdrawarc[thick, color=blue!70!black,tdplot_rotated_coords]{(A1)}{2.4}{-20}{20}{}{};
    
    \node[eqlabel, font=\large] (B1) at ($(V)!0.55!(RX3)+(0.0,-0.4,0.0)$) {$[\bm{\mathrm{b}}_q(\bm{\mathrm{R}})]_{n}$}; 
    \tdplotsetrotatedcoords{45}{180}{10};
    \tdplotdrawarc[thick, color=red!70!black,tdplot_rotated_coords]{($(V)!0.10!(RX3)+(0.0,-0.4,0.0)$)}{2}{-20}{20}{}{};
    \tdplotdrawarc[thick, color=red!70!black,tdplot_rotated_coords]{($(V)!0.10!(RX3)+(0.0,-0.4,0.0)$)}{3.2}{-20}{20}{}{};
    \tdplotdrawarc[thick, color=red!70!black,tdplot_rotated_coords]{($(V)!0.10!(RX3)+(0.0,-0.4,0.0)$)}{2.8}{-20}{20}{}{};
    \tdplotdrawarc[thick, color=red!70!black,tdplot_rotated_coords]{($(V)!0.10!(RX3)+(0.0,-0.4,0.0)$)}{2.4}{-20}{20}{}{};




    \begin{scope}[tdplot_screen_coords,shift={(-210pt,108pt)}]
      \node[user] at (12pt,108pt) {};
      \node[right, font=\large] at (25pt,108pt) {Users};

      \node[tx] at (12pt,88pt) {};
      \node[right, font=\large] at (25pt,88pt) {Tx movable antenna};

      \node[rx] at (12pt,68pt) {};
      \node[right, font=\large] at (25pt,68pt) {Rx movable antenna};

      \node[focus] at (12pt,48pt) {};
      \node[right, font=\large] at (25pt,48pt) {Imaging pixel};

      \draw[comm] (2pt,28pt) -- (22pt,28pt);
      \node[right, font=\large] at (25pt,28pt) {Communication link};

      \draw[radar] (2pt,8pt) -- (23pt,8pt);
      \node[right, font=\large] at (25pt,8pt) {Sensing link};

    \end{scope}

    \end{tikzpicture}
}
\vspace{-0.2cm}
\caption{Illustration of \ac{nf} \ac{isac} system with $K$ users and a \ac{2d} imaging \ac{roi}.}
\vspace{-0.4cm}
\label{fig:concept}
\end{figure}

We consider an \ac{ma}-enabled \ac{nf} \ac{isac} system, where a \ac{bs} simultaneously serves \(K\) single-antenna users and performs radar imaging over a \ac{2d} \ac{roi},
as illustrated in Fig.~\ref{fig:concept}. The \ac{bs} employs \(M_{\mathrm{t}}\) transmit and \(N_{\mathrm{r}}\) receive isotropic \acp{ma}, whose positions are adjustable within the planar aperture
\linebreaker{$
\mathcal{C} = \left\{ \mathbf{p} = [x,y,0]^\top \in \mathbb{R}^3 \;\middle|\; x,y \in \left[-\frac{D}{2},\frac{D}{2}\right] \right\}$,}{
\begin{equation}
    \mathcal{C} = \left\{ \mathbf{p} = [x,y,0]^\top \in \mathbb{R}^3 \;\middle|\; x,y \in \left[-\frac{D}{2},\frac{D}{2}\right] \right\},
\end{equation}} where $D$ denotes the aperture size. Self-interference from full-duplex radar operation is assumed to be perfectly canceled~\cite{abusanadJointImagingDownlink2025}. 
Both the users and the \ac{roi} lie in the radiating \ac{nf} region, characterized by $0.62\sqrt{D^3/\lambda}
\leq R \leq
2D^2 / \lambda$, where $R$ is the distance from the array aperture and \(\lambda=c_0/f_{\mathrm{c}}\) is the wavelength, with \(f_{\mathrm{c}}\) denoting the carrier frequency and \(c_0\) the speed of light \cite{quNearFieldIntegratedSensing2024}. Propagation follows spherical wavefronts under the Born approximation, such that the received field is linear in the scene reflectivity and multiple scattering is neglected \cite{jiangMASCOTMaskAssistedSpatial2025}. 
Furthermore, only \ac{los} communication paths and single-bounce target returns are considered.
The transmit and receive antenna positions are denoted by $\bm{\mathrm{t}}_m=[x^{\mathrm{t}}_{m},y^{\mathrm{t}}_{m},0]^\top\in\mathcal{C}$ and $\bm{\mathrm{r}}_n=[x^{\mathrm{r}}_{n},y^{\mathrm{r}}_{n},0]^\top\in\mathcal{C},$
and are collected in matrices
$
\bm{\mathrm{T}}
=[\bm{\mathrm{t}}_1,\ldots,\bm{\mathrm{t}}_{M_{\mathrm{t}}}]^\top \in \mathbb{R}^{M_{\mathrm{t}} \times 3}$
and $
\bm{\mathrm{R}}
=[\bm{\mathrm{r}}_1,\ldots,\bm{\mathrm{r}}_{N_{\mathrm{r}}}]^\top \in \mathbb{R}^{N_{\mathrm{r}}\times 3}.$
The antenna positions remain fixed during each observation interval comprising \(L\) symbol intervals. A minimum spacing of \(d_0=\lambda/2\) is imposed between all antenna elements to limit mutual coupling \cite{ahmedElectronicMicrowaveImaging}. The \ac{2d} \ac{roi} at depth $z_{\mathrm{ROI}}$ is discretized into \(Q\) \ac{los} pixels at positions
\(\bm{\mathrm{q}}_q=[x^{\mathrm{q}}_q,y^{\mathrm{q}}_q,z_{\mathrm{ROI}}]^\top\), \(q \in \left\{1,\ldots,Q\right\},\) while the users are excluded from the scene reflectivity. Each pixel has an unknown complex reflectivity \(\sigma_q\), collected in
$
\bm{\sigma}
=
[\sigma_1,\ldots,\sigma_Q]^\top
\in\mathbb{C}^{Q}.
$
The \(k\)-th communication user is located at
\(\bm{\mathrm{u}}_k=[x^{\mathrm{u}}_k,y^{\mathrm{u}}_k,z^{\mathrm{u}}_k]^\top\), \( k \in \left\{1,\ldots,K\right\}\).

\subsection{Communication and Imaging Signal Transmission}
The transmitted signal consists of \(K\) communication and \(S\) sensing streams sharing the same time--frequency resources \cite{rahalNearFieldWidebandISAC2025}. Assuming \(K\leq M_{\mathrm{t}}\), we set \(S=M_{\mathrm{t}}-K\) to exploit the remaining spatial \ac{dof} for sensing. 
Over \(L\) symbol intervals, the transmitted signal matrix is
\begin{equation}
\bm{\mathrm{X}}
= \bm{\mathrm{V}}^{\mathrm{co}}\bm{\mathrm{C}}
+ \bm{\mathrm{V}}^{\mathrm{ra}}\bm{\mathrm{S}}
\in \mathbb{C}^{M_{\mathrm{t}}\times L},
\label{eq:transmit_signal_matrix}
\end{equation}
where \(\bm{\mathrm{C}}=[\bm{\mathrm{c}}[1],\ldots,\bm{\mathrm{c}}[L]]\in\mathbb{C}^{K \times L}\)
and \(\bm{\mathrm{S}}=[\bm{\mathrm{s}}[1],\ldots,\bm{\mathrm{s}}[L]] \in\mathbb{C}^{S \times L}\)
are the communication and sensing symbol matrices, respectively,
with corresponding precoders
\(\bm{\mathrm{V}}^{\mathrm{co}}\in\mathbb{C}^{M_{\mathrm{t}}\times K}\)
and \(\bm{\mathrm{V}}^{\mathrm{ra}}\in\mathbb{C}^{M_{\mathrm{t}}\times S}\). Here, \(\bm{\mathrm{c}}[l]\) and \(\bm{\mathrm{s}}[l]\) denote the respective symbol vectors at interval \(l\). For compactness, the precoders and symbol matrices are combined as
\begin{equation}
    \bm{\mathrm{X}}
    =
    \overbrace{
    \begin{bmatrix}
        \bm{\mathrm{V}}^{\mathrm{co}} &
        \bm{\mathrm{V}}^{\mathrm{ra}}
    \end{bmatrix}
    }^{\bm{\mathrm{W}}}
    \overbrace{
    \begin{bmatrix}
        \bm{\mathrm{C}} \\
        \bm{\mathrm{S}}
    \end{bmatrix}
    }^{\bm{\mathrm{Z}}}
    =
    \bm{\mathrm{W}}\bm{\mathrm{Z}},
    \label{eq:compact_transmit_signal_matrix}
\end{equation}
where \(\bm{\mathrm{W}}\in\mathbb{C}^{M_{\mathrm{t}}\times(K+S)}\) and \(\bm{\mathrm{Z}}\in\mathbb{C}^{(K+S)\times L}\) denote the combined precoder and stacked symbol matrix, respectively. Assuming statistically independent communication and sensing sequences over sufficiently large \(L\), their sample covariance satisfies 
\begin{equation}
    \widetilde{\bm{\mathrm{R}}}_{zz}=\frac{1}{L}\bm{\mathrm{Z}} \bm{\mathrm{Z}}^H \approx \bm{\mathrm{I}}_{K+S}.
    \label{eq:total_symbol_cov_matrix}
\end{equation}
Consequently, the sample transmit covariance matrix becomes
\begin{equation}
\widetilde{\bm{\mathrm{R}}}_{xx}
= \frac{1}{L}\bm{\mathrm{X}}\bm{\mathrm{X}}^H
\approx \bm{\mathrm{W}}\bm{\mathrm{W}}^H.
\label{eq:transmit_signal_cov_matrix}
\end{equation}

\subsection{Near-Field Imaging Model}

The \ac{nf} transmit steering matrix can be defined as \(\bm{\mathrm{A}}
=
\left[
\bm{\mathrm{a}}_1(\bm{\mathrm{T}}), \ldots,
\bm{\mathrm{a}}_Q(\bm{\mathrm{T}})
\right]
\in \mathbb{C}^{M_{\mathrm{t}} \times Q}\) and the receive steering matrix as \(\bm{\mathrm{B}}=\left[\bm{\mathrm{b}}_1( \bm{\mathrm{R}}), \ldots,\bm{\mathrm{b}}_Q(\bm{\mathrm{R}})\right]\in \mathbb{C}^{N_{\mathrm{r}} \times Q},\)
with the respective elements given by
\begin{equation}
\left[\bm{\mathrm{a}}_{q}(\bm{\mathrm{T}})\right]_m
=
\frac{\lambda\exp\!\left(
-j \frac{2\pi f_{\mathrm{c}}}{c_0}
\|\bm{\mathrm{q}}_q-\bm{\mathrm{t}}_{m}\|
\right)}{4\pi \|\bm{\mathrm{q}}_q-\bm{\mathrm{t}}_{m}\|}, m \in \left\{1, \ldots, M_{\mathrm{t}}\right\},
\label{eq:transmit_imaging_channel_entry}
\end{equation} and \begin{equation}
\left[\bm{\mathrm{b}}_q(\bm{\mathrm{R}})\right]_{n}= \frac{\lambda\exp\!\left(- j \frac{2\pi f_{\mathrm{c}}}{c_0}\|\bm{\mathrm{q}}_q - \bm{\mathrm{r}}_{n}\|\right)}{4\pi \|\bm{\mathrm{q}}_q - \bm{\mathrm{r}}_{n}\|}, n \in \left\{1, \ldots, N_{\mathrm{r}}\right\}.
\label{eq:receive_imaging_channel_entry}
\end{equation}
To ensure sufficient illumination power throughout the \ac{roi}, the illumination power received at the $q$-th pixel, $P_q$, is constrained as follows
\begin{equation}
 P_q( \bm{\mathrm{T}}, \bm{\mathrm{W}}) =\bm{\mathrm{a}}_q^{\top}(\bm{\mathrm{T}})\bm{\mathrm{W}}\bm{\mathrm{W}}^{H} \bm{\mathrm{a}}^*_q(\bm{\mathrm{T}})  \geq P_{\mathrm{ROI}}, \forall q,
 \label{eq:ROI_power_constraint}
\end{equation}
where $P_{\mathrm{ROI}}$ denotes the minimum illumination power required at each pixel.
For the characterization of the \ac{psf}, we neglect noise and hardware impairments
\cite{abusanadJointImagingDownlink2025,jiangMASCOTMaskAssistedSpatial2025}. The received sensing signal is thus modeled as
\begin{equation}
    \bm{\mathrm{Y}}
    = \bm{\mathrm{B}}\operatorname{diag}(\bm{\sigma})
    \bm{\mathrm{A}}\mkern-4mu^{\top} \bm{\mathrm{X}}
    \in \mathbb{C}^{N_{\mathrm{r}} \times L}.
    \label{eq:received_imaging_signal_matrix}
\end{equation}


\subsection{PSF Formulation and Imaging Metrics}

Assuming perfect knowledge of the transmitted waveforms and \ac{nf} steering responses, correlating the received signal with the transmitted waveform yields
\begin{equation}
\frac{1}{L}\bm{\mathrm{Y}} \bm{\mathrm{X}}^H
\approx \bm{\widetilde{\mathrm{Y}}} =
\bm{\mathrm{B}} \operatorname{diag}(\bm{\sigma}) \bm{\mathrm{A}}\mkern-4mu^{\top} \widetilde{\bm{\mathrm{R}}}_{xx}.
\label{eq:received_imaging_signal_crosscorr}
\end{equation}
Vectorization gives the equivalent linear sensing model
\begin{equation}
\begin{aligned}
\tilde{\bm{\mathrm{y}}}
= \operatorname{vec}(\bm{\widetilde{\mathrm{Y}}})=
\underbrace{
\left(
\widetilde{\bm{\mathrm{R}}}_{xx}^{\top}
\otimes \bm{\mathrm{I}}_{N_{\mathrm{r}}}
\right)
\left(\bm{\mathrm{A}} \odot \bm{\mathrm{B}}\right)
}_{\bm{\mathrm{G}}}
\bm{\sigma},
\end{aligned}
\label{eq:vectorized_received_crosscorr_imaging_signal}
\end{equation}
where \(\bm{\mathrm{G}}\in\mathbb{C}^{M_{\mathrm{t}}N_{\mathrm{r}}\times Q}\)
is the effective sensing matrix. 
Thus, the transmit covariance $\widetilde{\bm{\mathrm{R}}}_{xx}$
shapes the imaging response. Matched-filter back-projection
yields \cite{jiangMASCOTMaskAssistedSpatial2025}
\begin{equation}
\hat{\bm{\sigma}} = \bm{\Phi}\tilde{\bm{\mathrm{y}}}
= \bm{\Phi}\bm{\mathrm{G}}\bm{\sigma},
\label{eq:linear_backprojection}
\end{equation}
where $\bm{\Phi}=[\bm{\phi}_1,\ldots,\bm{\phi}_Q]^{\top}
\in\mathbb{C}^{Q\times M_{\mathrm{t}}N_{\mathrm{r}}}$
is the amplitude-compensated \ac{nf} matched-filter
operator with elements
\linebreaker{\begin{equation}
\begin{aligned}
[\bm{\phi}_q]_p
=& \frac{16\pi^2}{\lambda^2}
\|\bm{\mathrm{q}}_q-\bm{\mathrm{r}}_n\|
\|\bm{\mathrm{q}}_q-\bm{\mathrm{t}}_m\|\\
&\times\exp\!\left(
j\frac{2\pi f_{\mathrm{c}}}{c_0}
\left(\|\bm{\mathrm{q}}_q-\bm{\mathrm{r}}_n\|
+\|\bm{\mathrm{q}}_q-\bm{\mathrm{t}}_m\|\right)
\right),
\end{aligned}
\end{equation}
}{
\begin{equation}
[\bm{\phi}_q]_p
= \frac{16\pi^2}{\lambda^2}
\|\bm{\mathrm{q}}_q-\bm{\mathrm{r}}_n\|
\|\bm{\mathrm{q}}_q-\bm{\mathrm{t}}_m\|
\times\exp\!\left(
j\frac{2\pi f_{\mathrm{c}}}{c_0}
\left(\|\bm{\mathrm{q}}_q-\bm{\mathrm{r}}_n\|
+\|\bm{\mathrm{q}}_q-\bm{\mathrm{t}}_m\|\right)
\right),
\end{equation}}
where $p=1,\ldots,M_{\mathrm{t}}N_{\mathrm{r}}$
indexes the transmit--receive antenna pairs $(m,n)$.
Accordingly, the \ac{psf} matrix is defined as
\begin{equation}
\bm{\mathrm{F}}(\bm{\mathrm{T}},\bm{\mathrm{R}},\bm{\mathrm{W}})
=
\bm{\Phi}\bm{\mathrm{G}}
\in \mathbb{C}^{Q \times Q}.
\label{eq:psf_matrix}
\end{equation}
For a unit scatterer at pixel \(q\), i.e.,
\(\bm{\sigma}=\bm{\mathrm{e}}_{q}\),
the corresponding \ac{psf} is the \(q\)-th column of
\(\bm{\mathrm{F}}\), denoted by $\bm{f}_{q} \in \mathbb{C}^{Q}$.

To assess the imaging performance in terms of the \ac{psf} sidelobe characteristics, we consider two complementary metrics: the \ac{isl}, which measures the total sidelobe energy and reflects potential reconstruction artifacts in \eqref{eq:linear_backprojection}, and the \ac{psl}, which captures the strongest sidelobe relative to the desired focal response, thereby indicating the risk for ghost target detection. For a unit scatterer at pixel $q$, the normalized \ac{isl} and \ac{psl} are respectively defined as
\linebreaker{
\begin{equation}
\mathrm{ISL}_{q}=
\frac{
\sum_{\substack{q'=1, q'\neq q}}^{Q}
\left|[\bm{\mathrm{f}}_{q}]_{q'}\right|^2
}{
\|
\bm{\mathrm{f}}_{q}\|^2
}, \mathrm{PSL}_{q}=
\frac{
\max\limits_{\substack{q'\in{1,\ldots,Q}, q'\neq q}}
\left|[\bm{\mathrm{f}}_{q}]_{q'}\right|^2
}{
\left|[\bm{\mathrm{f}}_{q}]_{q}\right|^2
}.
\end{equation}
}{
\begin{equation}
\mathrm{ISL}_{q}=
\frac{
\sum_{\substack{q'=1, q'\neq q}}^{Q}
\left|[\bm{\mathrm{f}}_{q}]_{q'}\right|^2
}{
\|
\bm{\mathrm{f}}_{q}\|^2
}, \qquad \mathrm{PSL}_{q}=
\frac{
\max\limits_{\substack{q'\in{1,\ldots,Q}, q'\neq q}}
\left|[\bm{\mathrm{f}}_{q}]_{q'}\right|^2
}{
\left|[\bm{\mathrm{f}}_{q}]_{q}\right|^2
}.
\end{equation}
}
Normalizing the \ac{isl} by the total \ac{psf} energy ensures scale independence and bounds it to $[0,1]$\footnote{Conventional normalization of the \ac{isl} by the mainlobe energy does not ensure values within $[0,1]$, complicating the balancing of objectives in weighted-sum optimization.}, whereas mainlobe normalization is sufficient for the \ac{psl}, since the true scatterer position corresponds to the maximum of the \ac{psf}. 
Finally, to account for all possible scatterer positions within the \ac{roi}, the metrics are averaged over all \(Q\) pixels as
\begin{equation}
\overline{\mathrm{ISL}}
=
\frac{1}{Q}
\sum_{q=1}^{Q}
\mathrm{ISL}_{q},
\qquad
\overline{\mathrm{PSL}}
=
\frac{1}{Q}
\sum_{q=1}^{Q}
\mathrm{PSL}_{q}.
\end{equation}

\subsection{Communication Model}

The \ac{nf} channel between the transmit array and user \(k\) is denoted by
\(\bm{\mathrm{h}}_k(\bm{\mathrm{T}})
=[h_{1,k},\ldots,h_{M_{\mathrm{t}},k}]^\top \in \mathbb{C}^{M_{\mathrm{t}}}\), with
\begin{equation}
[\bm{\mathrm{h}}_k(\bm{\mathrm{T}})]_{m} = \frac{\lambda\exp\!\left(- j \frac{2\pi f_{\mathrm{c}}}{c_0}\|\bm{\mathrm{u}}_k - \bm{\mathrm{t}}_{m}\|\right)}{4\pi \|\bm{\mathrm{u}}_k - \bm{\mathrm{t}}_{m}\|}, m \in \left\{1, \ldots, M_{\mathrm{t}}\right\}.
\label{eq:received_comm_channel_entry}
\end{equation} 
Using the transmit signal defined in \eqref{eq:transmit_signal_matrix}, the received signal at user \(k\) at symbol interval \(l\) is given by
\linebreaker{
\begin{equation}
\begin{aligned}
y_{c}^k[l]
={}&
\underbrace{
\bm{\mathrm{h}}_{k}^{\top}(\bm{\mathrm{T}})
\bm{\mathrm{w}}_{k} c_{k}[l]
}_{\text{desired signal}}
+
\underbrace{
\sum_{\substack{j=1, j\neq k}}^{K}
\bm{\mathrm{h}}_{k}^{\top}(\bm{\mathrm{T}})
\bm{\mathrm{w}}_{j} c_{j}[l]
}_{\text{multiuser interference}}
\\
&+
\underbrace{
\bm{\mathrm{h}}_{k}^{\top}(\bm{\mathrm{T}})
\bm{\mathrm{V}}^{\mathrm{radar}}
\bm{\mathrm{s}}[l]
}_{\text{sensing interference}}
+
n_{k}[l].
\end{aligned}
\label{eq:received_comm_signal_full}
\end{equation}
}{
\begin{equation}
y_{c}^k[l]
=
\underbrace{
\bm{\mathrm{h}}_{k}^{\top}(\bm{\mathrm{T}})
\bm{\mathrm{w}}_{k} c_{k}[l]
}_{\text{desired signal}}
+
\underbrace{
\sum_{\substack{j=1, j\neq k}}^{K}
\bm{\mathrm{h}}_{k}^{\top}(\bm{\mathrm{T}})
\bm{\mathrm{w}}_{j} c_{j}[l]
}_{\text{multiuser interference}}
+
\underbrace{
\bm{\mathrm{h}}_{k}^{\top}(\bm{\mathrm{T}})
\bm{\mathrm{V}}^{\mathrm{ra}}
\bm{\mathrm{s}}[l]
}_{\text{sensing interference}}
+
n_{k}[l].
\label{eq:received_comm_signal_full}
\end{equation}
}
where \(\bm{\mathrm{w}}_k \in \mathbb{C}^{M_{\mathrm{t}}}\) is the \(k\)-th column of the precoding matrix $\bm{\mathrm{W}}$ and
\(n_k[l]\sim\mathcal{CN}(0,\sigma_{\mathrm{n}}^2)\) denotes the \ac{awgn} with zero mean and variance $\sigma_{\mathrm{n}}^2.$ 
We assume that the sensing waveform is known a priori to the users, enabling perfect cancellation of the sensing interference \cite{huaOptimalTransmitBeamforming2023}. Hence, for unit-power and mutually independent communication symbols, the resulting \ac{sinr} becomes \cite{huaOptimalTransmitBeamforming2023, rahalNearFieldWidebandISAC2025} 
\begin{equation}
\mathrm{SINR}_{k}(\bm{\mathrm{T}},\bm{\mathrm{W}})=\frac{|\bm{\mathrm{h}}_{k}^{\top}(\bm{\mathrm{T}})\bm{\mathrm{w}}_{k}|^2
}{\sum_{j=1,j\neq k}^{K}|\bm{\mathrm{h}}_{k}^{\top}(\bm{\mathrm{T}})\bm{\mathrm{w}}_{j}|^2+\sigma_{\mathrm{n}}^2}.
\label{eq:sinr_k}
\end{equation}
The achievable rate of user \(k\) is therefore given by
\begin{equation}
\gamma_k(\bm{\mathrm{T}},\bm{\mathrm{W}}) = \log_2\left(1 + \mathrm{SINR}_{k}(\bm{\mathrm{T}},\bm{\mathrm{W}})\right).
\label{eq:communication_rate}
\end{equation}

\versiontext{
\section{Problem Formulation And Proposed Solution}
\subsection{Problem Formulation}
In this paper, we aim to minimize a weighted sum of \ac{isl} and \ac{psl} by jointly optimizing the transmit and receive \ac{ma} positions and the transmit beamforming matrix, subject to communication, sensing, and physical array constraints. The corresponding optimization problem is formulated as follows
\linebreaker{\begin{subequations}
	\label{eq:optimization_problem}
	\begin{align}
		&(\text{P})\quad
		\underset{\mathbf{T},\mathbf{R},\mathbf{W}}{\min}\quad
		 J_{0}(\mathbf{T},\mathbf{R},\mathbf{W}) =  w_{\mathrm{ISL}}\,\overline{\mathrm{ISL}} + w_{\mathrm{PSL}}\,\overline{\mathrm{PSL}} \notag\\
         &\text{s.t.}\notag &&\\
		& \gamma_k(\mathbf{T},\mathbf{W}) \ge \gamma_{\mathrm{min}}, \forall\, k, \label{eq:CommRate_constraint}&& \\
		& \min_{q} P_q(\mathbf{T}, \mathbf{W}) \ge P_{\mathrm{ROI}}, \label{eq:Pill_constraint}\\
		&  \mathrm{tr}(\mathbf{W}\mathbf{W}^H) \le P_{\mathrm{total}}, \label{eq:tot_transmit_power}\\
		& \|\mathbf{p}-\mathbf{p}'\|\geq d_0, \mathbf{p} \neq \mathbf{p}'\in \mathcal A, \label{eq:min_spacing_constraint} && \\
		& \mathbf{r}_n, \mathbf{t}_m \in \mathcal{C}, \forall n, \forall m.\label{eq:aperture_constraint}
	\end{align}
\end{subequations}
}{\begin{subequations}
	\label{eq:optimization_problem}
	\begin{align}
		(\text{P})\quad
		\underset{\mathbf{T},\mathbf{R},\mathbf{W}}{\min}\quad
		 &J_{0}(\mathbf{T},\mathbf{R},\mathbf{W}) =  w_{\mathrm{ISL}}\,\overline{\mathrm{ISL}} + w_{\mathrm{PSL}}\,\overline{\mathrm{PSL}} \notag\\
        \mathrm{s.t.}\quad 
        &\gamma_k(\mathbf{T},\mathbf{W}) \ge \gamma_{\mathrm{min}}, \forall\, k, \label{eq:CommRate_constraint}&& \\
		& \min_{q} P_q(\mathbf{T}, \mathbf{W}) \ge P_{\mathrm{ROI}}, \label{eq:Pill_constraint}\\
		&  \mathrm{tr}(\mathbf{W}\mathbf{W}^H) \le P_{\mathrm{total}}, \label{eq:tot_transmit_power}\\
		& \|\mathbf{p}-\mathbf{p}'\|\geq d_0, \mathbf{p} \neq \mathbf{p}'\in \mathcal A, \label{eq:min_spacing_constraint} && \\
		& \mathbf{r}_n, \mathbf{t}_m \in \mathcal{C}, \forall n, \forall m,\label{eq:aperture_constraint}
	\end{align}
\end{subequations}
}
where $\mathcal A =
\{\mathbf{r}_n\}_{n=1}^{N_{\mathrm{r}}}\cup
\{\mathbf{t}_m\}_{m=1}^{M_{\mathrm{t}}}$ and $w_{\mathrm{ISL}},w_{\mathrm{PSL}} \ge 0$ are the weighting coefficients for the \ac{isl} and \ac{psl}, respectively, such that $w_{\mathrm{ISL}} + w_{\mathrm{PSL}} = 1$. Constraint \eqref{eq:CommRate_constraint} guarantees that the downlink rate for each user remains above the threshold $\gamma_{\mathrm{min}}$, while \eqref{eq:Pill_constraint} enforces sufficient illumination throughout the ROI. Constraint \eqref{eq:tot_transmit_power} restricts the total transmit power to the available power budget $P_{\mathrm{total}}$. Finally, \eqref{eq:min_spacing_constraint} enforces the minimum inter-element spacing $d_0$ between all receive and transmit antennas, and \eqref{eq:aperture_constraint} confines all \acp{ma} to the feasible aperture $\mathcal{C}$.\par 

\subsection{Proposed Solution}
Problem $\mathrm{(P)}$ is challenging to solve due to (i) the coupling between antenna geometry and beamforming and the non-convexity of (ii) the \ac{psf}-based imaging objective, (iii) the communication-rate constraint \eqref{eq:CommRate_constraint}, and (iv) the minimum illumination power constraint \eqref{eq:Pill_constraint}. To tackle this problem, we propose a hybrid framework of \ac{ao} and metaheuristics that alternates between antenna-position optimization using \ac{pso} in subproblem $\mathrm{(P1)}$ and beamforming optimization using a \ac{ga} in subproblem $\mathrm{(P2)}$. Such approaches have proven effective for complex, high-dimensional optimization problems \cite{lahmeriUAVFormationResource2025a}. In the following, the proposed solution is described in detail.
\subsection{Movable Antenna Placement Optimization} 
\label{subsec:MA_optim}
In subproblem $\mathrm{(P1)}$, the transmit and receive \ac{ma} positions, $\mathbf{T}$ and $\mathbf{R}$, respectively, are optimized for a fixed beamforming matrix \(\bm{\mathrm{W}}\). The resulting antenna placement optimization subproblem is given by 
\begin{equation}
\begin{aligned}
(\mathrm{P}1)\quad
\underset{\bm{\mathrm{T}},\bm{\mathrm{R}}}{\min}\quad
& J_{0}(\bm{\mathrm{T}},\bm{\mathrm{R}},\bm{\mathrm{W}}) \\
\mathrm{s.t.}\quad
& \eqref{eq:Pill_constraint},
\eqref{eq:min_spacing_constraint},
\eqref{eq:aperture_constraint}.
\end{aligned}
\label{eq:objective_SP1}
\end{equation}
Subproblem $\mathrm{(P1)}$ remains challenging to solve due to its complex, nonlinear, and highly non-convex \ac{psf}-based objective function $J_0$. Motivated by its success in solving similar antenna topology optimization problems \cite{jiangMASCOTMaskAssistedSpatial2025, lahmeriUAVFormationResource2025a}, we adopt \ac{pso} to solve $\mathrm{(P1)}$. In preliminary experiments, we observed that enforcing \eqref{eq:CommRate_constraint} during the geometry update can lead to only marginal position changes across iterations. A similar observation was reported in \cite{maMovableAntennaEnhanced2026}. Therefore, we omit this constraint in $\mathrm{(P1)}$ for placement optimization in \eqref{eq:objective_SP1} and enforce it in $\mathrm{(P2)}$ for the subsequent beamforming update, see \eqref{eq:objective_SP2}, thereby ensuring the feasibility of the final solution. To optimize the transmit and receive \ac{ma} positions, we employ \ac{pso} with a population of $D_{\mathsf{PSO}}$ particles. Each particle encodes a candidate antenna geometry $(\bm{\mathrm{T}}_{i}^{(t)},\bm{\mathrm{R}}_{i}^{(t)})$ by specifying the $x$- and $y$-coordinates of all transmit and receive antennas. In particular, the position of particle $i$ at iteration $t$ is given by
\linebreaker{\begin{equation}
\begin{aligned}
\mathbf p_i^{(t)} =
&\big[x^{\mathrm{t}}_{1}, y^{\mathrm{t}}_{1},\ldots,
x^{\mathrm{t}}_{M_{\mathrm t}}, y^{\mathrm{t}}_{M_{\mathrm t}},\\
&\quad x^{\mathrm{r}}_{1}, y^{\mathrm{r}}_{1},\ldots,
x^{\mathrm{r}}_{N_{\mathrm r}}, y^{\mathrm{r}}_{N_{\mathrm r}}
\big]^{\top}\in\mathbb R^{2(M_{\mathrm t}+N_{\mathrm r})}.
\end{aligned}
\label{eq:pso_particle}
\end{equation}}{\begin{equation}
\mathbf p_i^{(t)}
=
\left[ x^{\mathrm{t}}_{1}, y^{\mathrm{t}}_{1},\ldots,
x^{\mathrm{t}}_{{M_{\mathrm t}}}, y^{\mathrm{t}}_{{M_{\mathrm t}}},
x^{\mathrm{r}}_{1}, y^{\mathrm{r}}_{1},\ldots,
x^{\mathrm{r}}_{{N_{\mathrm r}}}, y^{\mathrm{r}}_{{N_{\mathrm r}}}
\right]^{\top}
\in\mathbb R^{2(M_{\mathrm t}+N_{\mathrm r})}.
\label{eq:pso_particle}
\end{equation}}
Each particle is associated with a velocity vector $\mathbf v_i^{(t)}$ of the same dimension, which determines its displacement within the search space. Following the standard \ac{pso} formulation \cite{kennedyParticleSwarmOptimization1995}, particle velocities and positions are updated as follows
\linebreaker{\begin{align}
\mathbf v_i^{(t+1)}
&=
\omega^{(t)}\mathbf v_i^{(t)}
+c_{1}\bm{\varphi}_{1,i}\bullet\left(\mathbf p_{\mathrm{best},i}^{(t)}-\mathbf p_i^{(t)}\right) \nonumber\\
&\quad+
c_{2}\bm{\varphi}_{2,i}
\bullet\left(\mathbf g_{\mathrm{best}}^{(t)}-\mathbf p_i^{(t)}\right),
\label{eq:PSO_velocity_update}\\
\mathbf p_i^{(t+1)}
&=
\mathbf p_i^{(t)}+\mathbf v_i^{(t+1)},
\label{eq:PSO_position_update}
\end{align}}{\begin{align}
\mathbf v_i^{(t+1)}
&=
\omega^{(t)}\mathbf v_i^{(t)}
+c_{1}\bm{\varphi}_{1,i}\bullet\left(\mathbf p_{\mathrm{best},i}^{(t)}-\mathbf p_i^{(t)}\right)+
c_{2}\bm{\varphi}_{2,i}
\bullet\left(\mathbf g_{\mathrm{best}}^{(t)}-\mathbf p_i^{(t)}\right),
\label{eq:PSO_velocity_update}\\
\mathbf p_i^{(t+1)}
&=
\mathbf p_i^{(t)}+\mathbf v_i^{(t+1)},
\label{eq:PSO_position_update}
\end{align}}
where $\mathbf p_{\mathrm{best},i}^{(t)}$ and $\mathbf g_{\mathrm{best}}^{(t)}$ are the current personal and global best-known positions, respectively. The coefficients $c_1$ and $c_2$ are the cognitive and social learning factors, respectively, $\bm{\varphi}_{1,i}$ and $\bm{\varphi}_{2,i}$ are independent uniform random vectors in $[0,1]^{2(M_{\mathrm t}+N_{\mathrm r})}$. The inertia weight $\omega^{(t)}$ decreases linearly to balance exploration and exploitation
\begin{equation}
\omega^{(t)}=\omega_{\max}-\frac{t}{T_{\mathrm{PSO}}}(\omega_{\max}-\omega_{\min}),  t=0,\ldots,T_{\mathrm{PSO}},
\label{eq:inertia_weight}
\end{equation} where $\omega_{\max}$ and $\omega_{\min}$ are the start and end inertia weights, respectively, and $T_{\mathrm{PSO}}$ is the maximum number of \ac{pso} iterations. By defining a reflective wall \cite{robinsonParticleSwarmOptimization2004} based on constraint \eqref{eq:aperture_constraint}, particles crossing the aperture boundary are reflected back into $\mathcal C$ with the sign of the corresponding velocity components changed. The illumination and minimum-spacing constraints are handled using Deb's feasibility rules \cite{debEAcientConstraintHandling}, which prioritize feasible solutions and rank infeasible ones by their constraint violations. Accordingly, for fixed $\bm{\mathrm{W}}$, the fitness is defined as
\linebreaker{\begin{equation}
\begin{aligned}
&J_1(\bm{\mathrm{p}}_i^{(t)}) \\
&=\begin{cases}
J_0(\bm{\mathrm{T}}_{i}^{(t)},\bm{\mathrm{R}}_{i}^{(t)},\bm{\mathrm{W}}),
& \text{if} \ \bm{\mathrm{p}}_i^{(t)}\in\mathcal F_1,\\
1+e_\mathrm{b}(\bm{\mathrm{T}}_{i}^{(t)}, \bm{\mathrm{W}})
+ e_\mathrm{d}(\bm{\mathrm{T}}_{i}^{(t)},\bm{\mathrm{R}}_{i}^{(t)}),
& \text{otherwise},
\end{cases}
\label{eq:pso_fitness}
\end{aligned}
\end{equation}}{\begin{equation}
J_1(\bm{\mathrm{p}}_i^{(t)}) =\begin{cases}
J_0(\bm{\mathrm{T}}_{i}^{(t)},\bm{\mathrm{R}}_{i}^{(t)},\bm{\mathrm{W}}),
& \text{if} \ \bm{\mathrm{p}}_i^{(t)}\in\mathcal F_1,\\
1+e_\mathrm{b}(\bm{\mathrm{T}}_{i}^{(t)}, \bm{\mathrm{W}})
+ e_\mathrm{d}(\bm{\mathrm{T}}_{i}^{(t)},\bm{\mathrm{R}}_{i}^{(t)}),
& \text{otherwise},
\end{cases}
\label{eq:pso_fitness}
\end{equation}}
where $\mathcal F_1$ is the feasible set of $(\mathrm{P}1)$ defined by \eqref{eq:Pill_constraint}, \eqref{eq:min_spacing_constraint}, and \eqref{eq:aperture_constraint}. Since the objective is upper-bounded by one, the fixed penalty offset is set to $1$ to ensure consistent fitness comparisons across iterations while prioritizing feasible solutions. The constraint violations are quantified as
\begin{align}
e_\mathrm{b}(\bm{\mathrm{T}}, \bm{\mathrm{W}})
&=\left[P_{\mathrm{ROI}}
-\min_q P_q(\bm{\mathrm{T}},\bm{\mathrm{W}})\right]_+,
\label{eq:penalty_illumination_power}
\\
e_\mathrm{d}(\bm{\mathrm{T}}, \bm{\mathrm{R}})
&=\sum_{\substack{\mathbf p,\mathbf p'\in\mathcal A\\
\mathbf p\neq\mathbf p'}}
\left[d_0-\|\mathbf p-\mathbf p'\|\right]_+.
\end{align}
The proposed procedure to solve subproblem $(\mathrm{P}1)$ is summarized in \textbf{Algorithm}~\ref{alg:pso}.
\begin{algorithm}[ht]
\caption{\ac{pso}-Based Antenna-Position Optimization}
\label{alg:pso}
\begin{algorithmic}[1]
\Require $\bm{\mathrm T},\bm{\mathrm R},\bm{\mathrm W}$,
$D_{\mathrm{PSO}}$, $T_{\mathrm{PSO}}$
\State Initialize swarm of $D_{\mathrm{PSO}}$ particles $(\bm{\mathrm p}^{(0)}_i, \ldots, \bm{\mathrm p}^{(0)}_{D_{\mathrm{PSO}}})$ with $\bm{\mathrm p}^{(0)}_1 \gets(\bm{\mathrm T},\bm{\mathrm R})$, initial random velocities $(\bm{\mathrm v}^{(0)}_i, \allowbreak \ldots,  \allowbreak\bm{\mathrm v}^{(0)}_{D_{\mathrm{PSO}}})$ and inertia weight $\omega^{(0)}$ using \eqref{eq:inertia_weight}
\State Evaluate particles fitness using \eqref{eq:pso_fitness}
\State Initialize personal and global best positions
\For{$t=1,\ldots,T_{\mathrm{PSO}}$}
    \For{each particle $i$}
        \State Update velocity and position using
        \eqref{eq:PSO_velocity_update}--\eqref{eq:PSO_position_update}
        \State Apply aperture boundary reflection 
        \State Evaluate fitness using \eqref{eq:pso_fitness}
        \State Update personal best position
    \EndFor
    \State Update global best position
    \State Update inertia weight $\omega^{(t)}$ based on \eqref{eq:inertia_weight}
\EndFor
\State \Return $(\bm{\mathrm T},\bm{\mathrm R})
\gets \bm{\mathrm g}^{(T_{\mathrm{PSO}})}_{\mathrm{best}}$
\end{algorithmic}
\end{algorithm}

\subsection{Beamforming Update by GA}
For a fixed antenna geometry, the transmit beamforming matrix is optimized by solving
\begin{equation}
\begin{aligned}
(\mathrm{P}2)\quad
\underset{\bm{\mathrm{W}}}{\min}\quad
& J_{0}(\bm{\mathrm{T}},\bm{\mathrm{R}},\bm{\mathrm{W}}) \\
\mathrm{s.t.}\quad
& \eqref{eq:CommRate_constraint},
\eqref{eq:Pill_constraint},
\eqref{eq:tot_transmit_power}.
\end{aligned}
\label{eq:objective_SP2}
\end{equation}
Due to the high dimensionality and non-convexity of the objective function and constraints, we employ a continuous \ac{ga}~\cite{hauptPracticalGeneticAlgorithms2004} to solve $(\mathrm{P}2)$. The \ac{ga} maintains a population of $D_{\mathsf{GA}}$ chromosomes, each encoding the magnitudes and phases of the complex beamforming coefficients. Specifically, chromosome $i$ at iteration $t$ is represented by
\begin{equation}
\bm{\mathrm{g}}_i^{(t)}
=
\begin{bmatrix}
(\bm{\mathrm{m}}_i^{(t)})^{\top}, &
(\bm{\phi}_i^{(t)})^{\top}
\end{bmatrix}^{\top}
\in [0,1]^{2M_{\mathrm t}(K+S)},
\label{eq:chromosome}
\end{equation}
where $\bm{\mathrm{m}}_i^{(t)},\bm{\phi}_i^{(t)}
\in[0,1]^{M_{\mathrm t}(K+S)}$ contain the normalized magnitudes and phases, respectively. The initial population is generated by uniformly sampling the chromosome entries over \([0,1]\), while a previously obtained beamformer, if available, is inserted as a warm start. To evaluate each chromosome, its magnitude and phase vectors are first reshaped into matrices of dimension $M_{\mathrm t}\times(K+S)$. The corresponding complex beamforming matrix is reconstructed through element-wise magnitude and phase mapping as
\begin{equation}
\bm{\mathrm{W}}_i^{(t)}
=
\bm{\mathrm{m}}_i^{(t)}
\bullet
\exp\left(j2\pi\bm{\phi}_i^{(t)}\right).
\label{eq:chromosome_to_beamformer}
\end{equation}
To satisfy the total transmit-power constraint, the resulting beamformer is normalized according to
\begin{equation}
\bm{\mathrm{W}}_i^{(t)}
\leftarrow
\sqrt{\frac{P_{\mathsf{total}}}
{\|\bm{\mathrm{W}}_i^{(t)}\|_{\mathrm{F}}^2}}
\bm{\mathrm{W}}_i^{(t)}.
\label{eq:beamformer_normalization}
\end{equation}
This normalization fully utilizes the transmit-power budget without loss of optimality, since a common scaling of $\mathbf W$ leaves the normalized \ac{isl} and \ac{psl} unchanged, while increasing the illumination power and not decreasing the user \acp{sinr}. Hence, the search can be restricted to full-power beamformers. For compact notation, we denote by $\operatorname{encode}(\mathbf W)$ the mapping of a complex beamforming matrix to the chromosome representation in \eqref{eq:chromosome}, obtained from its normalized magnitudes and phases. Conversely, $\operatorname{decode}(\mathbf g)$ denotes the reconstruction of the corresponding complex beamforming matrix according to \eqref{eq:chromosome_to_beamformer}, followed by the power normalization in \eqref{eq:beamformer_normalization}. Over $T_{\mathsf{GA}}$ generations, the population evolves through tournament selection, arithmetic crossover, and mutation, following standard \ac{ga} principles \cite{hauptPracticalGeneticAlgorithms2004}. Crossover generates weighted combinations of parent chromosomes using a common blending coefficient $\beta$ uniformly distributed in $[0,1]$. Mutation randomly replaces a fraction $\mu$ of the entries across all chromosomes except the best-ranked one with independent samples drawn uniformly from $[0,1]$. Thus, $\mu(D_{\mathrm{GA}}-1)2M_{\mathrm t}(K+S)$ entries are randomly selected for mutation. Elitism retains the best $D_{\mathrm{kept}}$ chromosomes per generation. To reduce computational cost, only new or modified chromosomes are evaluated. Constraints are handled as in the antenna-position optimization, prioritizing feasible solutions and ranking infeasible ones by their constraint violations \cite{debEAcientConstraintHandling}. Unlike the \ac{pso} update, the \ac{ga} enforces the communication-rate constraints, allowing feasibility to be restored after antenna-position updates. Accordingly, the fitness is defined as
\linebreaker{\begin{equation}
\begin{aligned}
    &J_2(\bm{\mathrm{g}}_i^{(t)}) \\
&=\begin{cases}
J_0(\bm{\mathrm{T}},\bm{\mathrm{R}},
\bm{\mathrm{W}}_i^{(t)}),
& \text{if} \ \bm{\mathrm{g}}_i^{(t)}\in\mathcal F_2,\\
J_{2}^{\max}+e_\mathrm{a}(\bm{\mathrm{T}},\bm{\mathrm{W}}_i^{(t)}) + e_\mathrm{b}(\bm{\mathrm{T}},\bm{\mathrm{W}}_i^{(t)}),
& \text{otherwise},
\end{cases}
\label{eq:ga_fitness}
\end{aligned}
\end{equation}}{\begin{equation}
J_2(\bm{\mathrm{g}}_i^{(t)}) =
\begin{cases}
J_0(\bm{\mathrm{T}},\bm{\mathrm{R}},
\bm{\mathrm{W}}_i^{(t)}),
& \text{if} \ \bm{\mathrm{g}}_i^{(t)}\in\mathcal F_2,\\
J_{2}^{\max}+e_\mathrm{a}(\bm{\mathrm{T}},\bm{\mathrm{W}}_i^{(t)}) + e_\mathrm{b}(\bm{\mathrm{T}},\bm{\mathrm{W}}_i^{(t)}),
& \text{otherwise},
\end{cases}
\label{eq:ga_fitness}
\end{equation}}
where $\mathcal F_2$ is the feasible set of
$(\mathrm{P}2)$ defined by
\eqref{eq:CommRate_constraint},
\eqref{eq:Pill_constraint}, and
\eqref{eq:tot_transmit_power}. Here, $J_{2}^{\max}$
is the largest feasible objective value in the
current population, set to $1$ if none exists.
Since normalization ensures the transmit-power
constraint, the remaining violations are the illumination power, see \eqref{eq:penalty_illumination_power}, and communication rate violation given by
\begin{align}
e_\mathrm{a}(\bm{\mathrm{T}},\bm{\mathrm{W}})=\sum_{k=1}^{K}
\left[\gamma_{\mathrm{min}}
-\gamma_k(\bm{\mathrm{T}},\bm{\mathrm{W}})\right]_+.
\end{align}
The proposed procedure to solve subproblem $(\mathrm{P}2)$ is summarized in \textbf{Algorithm}~\ref{alg:ga}.
\begin{algorithm}[ht]
\caption{\ac{ga}-Based Beamforming Optimization}
\label{alg:ga}
\begin{algorithmic}[1]
\Require $\bm{\mathrm T},\bm{\mathrm R},\bm{\mathrm W}$,
$D_{\mathrm{GA}}$, $T_{\mathrm{GA}}$
\State Initialize population of $D_{\mathrm{GA}}$ chromosomes $(\bm{\mathrm g}^{(0)}_1, \allowbreak \ldots, \allowbreak \bm{\mathrm g}^{(0)}_{D_{\mathrm{GA}}})$ with $\bm{\mathrm g}^{(0)}_1 \gets\operatorname{encode}(\bm{\mathrm W})$ 
\State Evaluate chromosomes fitness using \eqref{eq:ga_fitness}
\State Rank chromosomes by fitness
\For{$t=1,\ldots,T_{\mathrm{GA}}$}
    \State Retain $D_{\mathrm{kept}}$ elite chromosomes
    \While{population size $< D_{\mathrm{GA}}$}
        \State Select parents via tournament selection
        \State Generate offspring via arithmetic crossover
        \State Add offspring to population
    \EndWhile
    \State Mutate entries at rate $\mu$, except $\bm{\mathrm g}^{(t)}_1$
    \State Evaluate changed chromosomes fitness using \eqref{eq:ga_fitness}
    \State Rank chromosomes by fitness
\EndFor
\State \Return $\bm{\mathrm W}
\gets \operatorname{decode}(\bm{\mathrm g}^{(T_{\mathrm{GA}})}_1)$
\end{algorithmic}
\end{algorithm}
\vspace{-0.3cm}

\subsection{Overall AO Algorithm}
\label{subsec:alternating_scheme}
The proposed \ac{ao} algorithm combines the \ac{pso}-based \ac{ma} update and the \ac{ga}-based beamforming update within an outer iterative loop. Preliminary simulations indicated that fully solving both subproblems at every outer iteration incurs high computational cost and restricts the exploration of the joint solution space due to the strong coupling between antenna placement and beamforming. Therefore, we propose an adaptive \ac{ao} framework that dynamically controls the number of inner iterations to balance exploration and exploitation. In each \ac{ao} iteration $\alpha$, we maintain a \emph{current solution}
$\widetilde{\mathcal{X}} \triangleq
\{\widetilde{\mathbf{T}},\widetilde{\mathbf{R}},\widetilde{\mathbf{W}}\}$,
which initializes the subsequent updates, and a \emph{best solution}
$\mathcal{X}_{\star}$, which stores the best feasible solution found so far. Maintaining these two solutions allows the stochastic optimizers to explore alternative designs without losing the best feasible solution. At each outer iteration $\alpha$, the \ac{pso} first updates the antenna positions based on $\widetilde{\mathcal{X}}$, yielding a candidate geometry $(\mathbf T,\mathbf R)$. Since the \ac{pso} update relaxes the communication-rate constraints, the resulting geometry may render the current beamformer infeasible. Subsequently, the \ac{ga} optimizer refines the beamformer $\mathbf{W}$ to improve the sensing objective while satisfying the feasible set of problem $\mathrm{(P)}$, denoted by $\mathcal F$. However, a single \ac{ga} update may be insufficient to adapt the beamformer to a substantially changed antenna geometry, potentially limiting the exploration of the joint solution space. Therefore, up to $\eta$ consecutive \ac{ga} calls are performed for beamforming refinement, terminating early once a feasible candidate improving the best-known objective is obtained. To adapt the computational effort to the evolution of the antenna geometry, the maximum number of \ac{ga} updates $\eta$ is determined using a counter $c$, which records the number of consecutive outer iterations with negligible geometry changes relative to the current solution $\widetilde{\mathcal{X}}$ (see Line~\ref{alg_line:dgeo} of \textbf{Algorithm}~\ref{alg:alternating_optimization}).When a significant geometry change occurs, the number of \ac{ga} updates is set to
\begin{equation}
\eta=\min\left(\eta_{\mathrm{max}},
\max\left(\eta_{\mathrm{min}},c\right)\right),
\end{equation}
where $\eta_{\mathrm{min}}$ and $\eta_{\mathrm{max}}$ denote minimum and maximum number of beamforming refinements, respectively, after which the counter $c$ is reset to $0$. Consequently, a significant geometry change following several iterations with negligible displacement permits more beamforming updates to accommodate the new antenna configuration and restore feasibility. The bounds $\eta_{\mathsf{min}}$ and $\eta_{\mathsf{max}}$ limit the refinement effort, while unchanged geometries receive only a single \ac{ga} update, thereby reducing the computational cost. After the beamforming stage, the resulting candidate solution $\mathcal X=(\mathbf T,\mathbf R,\mathbf W)$ is evaluated with respect to the feasible set $\mathcal F$ of problem $\mathrm{(P)}$. If $\mathcal X\notin\mathcal F$, the candidate is rejected and the current solution remains unchanged. Otherwise, the best solution is updated whenever the candidate achieves a lower objective value, i.e.,
\begin{equation}
\mathcal X_\star\gets\mathcal X
\quad\text{if}\quad
J_0(\mathcal X)<J_0(\mathcal X_\star).
\end{equation}
However, a feasible candidate does not necessarily replace the current solution. To allow controlled exploration, the candidate is accepted as the new current solution only if
\begin{equation}
J_0(\mathcal X)
\leq
J_0(\widetilde{\mathcal X})+\Delta_{\mathsf{acc}},
\label{eq:acceptance_rule}
\end{equation}
where $\Delta_{\mathsf{acc}}$ is an acceptance margin. If \eqref{eq:acceptance_rule} is satisfied, $\widetilde{\mathcal X}\gets\mathcal X$, otherwise, the previous current solution is retained. This mechanism permits slightly inferior feasible solutions to be accepted, thereby facilitating exploration and reducing the risk of premature stagnation. Meanwhile, the best solution is updated only upon improvement, ensuring that its objective value is monotonically non-increasing. After $\alpha_{\mathsf{AO}}$ outer iterations, the algorithm returns $\mathcal X_\star$ as the final solution. 
The overall procedure is summarized in \textbf{Algorithm}~\ref{alg:alternating_optimization}.
\linebreaker{\versiontext{\begin{algorithm}[ht]
\caption{Proposed \ac{ao}-based Solution}
\label{alg:alternating_optimization}
\begin{algorithmic}[1]
\State Initialize $\mathcal X_\star= \widetilde{\mathcal X}
\gets(\widetilde{\mathbf T},\widetilde{\mathbf R},\widetilde{\mathbf W})$, $c\gets0$, $\alpha_{\mathsf{AO}}$, $\eta_{\mathrm{min}}$, $\eta_{\mathrm{max}}$, and $\Delta_{\mathrm{acc}}$
\For{$\alpha=1,\ldots,\alpha_{\mathsf{AO}}$}
    \State  $r\gets0$
    \State Obtain $(\mathbf T,\mathbf R)$ using \textbf{Algorithm}~\ref{alg:pso} with $(\widetilde{\mathbf T},\widetilde{\mathbf R},\widetilde{\mathbf W})$
    \State $d_{\mathrm{geo}}\gets
           \|[\mathbf{T}^{\top},\mathbf R^{\top}]^{\top}-
             [\widetilde{\mathbf T}^{\top},
              \widetilde{\mathbf R}^{\top}]^{\top}\|_{\mathrm F}$ \label{alg_line:dgeo}

    \If{$d_{\mathrm{geo}}\leq10^{-8}$}
        \State $c\gets c+1,\quad\eta\gets1$
    \Else
        \State $\eta
        \gets \min\!\left(\eta_{\mathrm{max}},\max\!\left(\eta_{\mathrm{min}},c\right)\right),
               \quad c\gets0$
    \EndIf
    \State $\mathbf W \gets\widetilde{\mathbf W}$
\Repeat
    \State $r\gets r+1$
    \State Obtain $\mathbf W$ using \textbf{Algorithm}~\ref{alg:ga} with $(\mathbf T,\mathbf R,\mathbf W)$ 
    \State $\mathcal X\gets(\mathbf T,\mathbf R,\mathbf W)$
\Until{$r=\eta$ or
    $(\mathcal X\in\mathcal F$ and
    $J_0(\mathcal X)<J_0(\mathcal X_\star))$}
\If{$\mathcal X\in\mathcal F$ and
    $J_0(\mathcal X)<J_0(\mathcal X^\star)$}
    \State $\mathcal X^\star\gets\mathcal X$
\EndIf

\If{$\mathcal X\in\mathcal F$ and
    $J_0(\mathcal X)\le J_0(\widetilde{\mathcal X})+\Delta_{\rm acc}$}
    \State $\widetilde{\mathcal X}\gets\mathcal X$
\EndIf
\EndFor
\State \Return $\mathcal X_\star$
\end{algorithmic}
\end{algorithm}}{
\begin{algorithm}[ht] 
\caption{Proposed \ac{ao}-based Solution}
\label{alg:alternating_optimization}
\begin{algorithmic}[1]
\State Initialize $\mathcal X_\star= \widetilde{\mathcal X}
\gets(\widetilde{\mathbf T},\widetilde{\mathbf R},\widetilde{\mathbf W})$, $c\gets0$, $\alpha_{\mathsf{AO}}$, $\eta_{\mathrm{min}}$, $\eta_{\mathrm{max}}$, and $\Delta_{\mathrm{acc}}$
\For{$\alpha=1,\ldots,\alpha_{\mathsf{AO}}$}
    \State $(\mathbf T,\mathbf R)\gets
           \operatorname{PSO}(\widetilde{\mathbf T},
                              \widetilde{\mathbf R},
                              \widetilde{\mathbf W}),$ $\quad r\gets0,$  $\quad\mathbf W\gets\widetilde{\mathbf W}$
    \State $d_{\mathrm{geo}}\gets
           \|[\mathbf{T}^{\top},\mathbf R^{\top}]^{\top}-
             [\widetilde{\mathbf T}^{\top},
              \widetilde{\mathbf R}^{\top}]^{\top}\|_{\mathrm F}$ \label{alg_line:dgeo}

    \If{$d_{\mathrm{geo}}\leq10^{-8}$}
        \State $c\gets c+1,\quad\eta\gets1$
    \Else
        \State $\eta
        \gets \min\!\left(\eta_{\mathrm{max}},\max\!\left(\eta_{\mathrm{min}},c\right)\right),
               \quad c\gets0$
    \EndIf
\Repeat
    \State $r\gets r+1$
    \State $\mathbf W\gets\operatorname{GA}(\mathbf T,\mathbf R,\mathbf W)$,
               $\quad\mathcal X\gets(\mathbf T,\mathbf R,\mathbf W)$
\Until{$r=\eta$ or
    $(\mathcal X\in\mathcal F$ and
    $J_0(\mathcal X)<J_0(\mathcal X_\star))$}
\If{$\mathcal X\in\mathcal F$ and
    $J_0(\mathcal X)<J_0(\mathcal X^\star)$}
    \State $\mathcal X^\star\gets\mathcal X$
\EndIf
\If{$\mathcal X\in\mathcal F$ and
    $J_0(\mathcal X)\le J_0(\widetilde{\mathcal X})+\Delta_{\rm acc}$}
    \State $\widetilde{\mathcal X}\gets\mathcal X$
\EndIf
\EndFor
\State \Return $\mathcal X_\star$
\end{algorithmic}
\end{algorithm}}
}{}
Its has a polynomial worst-case computational complexity given by
\linebreaker{
\begin{equation}
\begin{aligned}
\mathcal{O}\Bigl(
&\alpha_{\mathsf{AO}}
\Bigl[
D_{\mathsf{PSO}}(T_{\mathsf{PSO}}+1)
+\eta_{\mathsf{max}} D_{\mathsf{GA}}(T_{\mathsf{GA}}+1)
\Bigr] \\
&\times
\Bigl[
QM_{\mathrm{t}}^{2}N_{\mathrm{r}}^{2}
+Q^{2}M_{\mathrm{t}}N_{\mathrm{r}}
\Bigr]
\Bigr).
\end{aligned}
\end{equation}
}{
\begin{equation}
\mathcal{O}\Bigl(
\alpha_{\mathsf{AO}}
\Bigl[
D_{\mathsf{PSO}}(T_{\mathsf{PSO}}+1)
+\eta_{\mathsf{max}}D_{\mathsf{GA}}(T_{\mathsf{GA}}+1)
\Bigr] \times
\Bigl[
QM_{\mathrm{t}}^{2}N_{\mathrm{r}}^{2}
+Q^{2}M_{\mathrm{t}}N_{\mathrm{r}}
\Bigr]
\Bigr).
\end{equation}
}
}
{
\section{Problem Formulation And Proposed Solution}
In this paper, we aim to minimize a weighted sum of \ac{isl} and \ac{psl} by jointly optimizing the transmit and receive \ac{ma} positions and the transmit beamforming matrix, subject to communication, sensing, and physical array constraints. The corresponding optimization problem is formulated as follows
\linebreaker{\begin{subequations}
	\label{eq:optimization_problem}
	\begin{align}
		&(\text{P})\quad
		\underset{\mathbf{T},\mathbf{R},\mathbf{W}}{\min}\quad
		 J_{0}(\mathbf{T},\mathbf{R},\mathbf{W}) =  w_{\mathrm{ISL}}\,\overline{\mathrm{ISL}} + w_{\mathrm{PSL}}\,\overline{\mathrm{PSL}} \notag\\
         &\text{s.t.}\notag &&\\
		& \gamma_k(\mathbf{T},\mathbf{W}) \ge \gamma_{\mathrm{min}}, \forall\, k, \label{eq:CommRate_constraint}&& \\
		& \min_{q} P_q(\mathbf{T}, \mathbf{W}) \ge P_{\mathrm{ROI}}, \label{eq:Pill_constraint}\\
		&  \mathrm{tr}(\mathbf{W}\mathbf{W}^H) \le P_{\mathrm{total}}, \label{eq:tot_transmit_power}\\
		& \|\mathbf{p}-\mathbf{p}'\|\geq d_0, \mathbf{p} \neq \mathbf{p}'\in
\{\mathbf{r}_n\}_{n=1}^{N_{\mathrm{r}}}\cup
\{\mathbf{t}_m\}_{m=1}^{M_{\mathrm{t}}}, \label{eq:min_spacing_constraint} && \\
		& \mathbf{r}_n, \mathbf{t}_m \in \mathcal{C}, \forall n, \forall m.\label{eq:aperture_constraint}
	\end{align}
\end{subequations}
}{\begin{subequations}
	\label{eq:optimization_problem}
	\begin{align}
		(\text{P})\quad
		\underset{\mathbf{T},\mathbf{R},\mathbf{W}}{\min}\quad
		 &J_{0}(\mathbf{T},\mathbf{R},\mathbf{W}) =  w_{\mathrm{ISL}}\,\overline{\mathrm{ISL}} + w_{\mathrm{PSL}}\,\overline{\mathrm{PSL}} \notag\\
        \mathrm{s.t.}\quad 
        &\gamma_k(\mathbf{T},\mathbf{W}) \ge \gamma_{\mathrm{min}}, \forall\, k, \label{eq:CommRate_constraint}&& \\
		& \min_{q} P_q(\mathbf{T}, \mathbf{W}) \ge P_{\mathrm{ROI}}, \label{eq:Pill_constraint}\\
		&  \mathrm{tr}(\mathbf{W}\mathbf{W}^H) \le P_{\mathrm{total}}, \label{eq:tot_transmit_power}\\
		& \|\mathbf{p}-\mathbf{p}'\|\geq d_0, \mathbf{p} \neq \mathbf{p}'\in
\{\mathbf{r}_n\}_{n=1}^{N_{\mathrm{r}}}\cup
\{\mathbf{t}_m\}_{m=1}^{M_{\mathrm{t}}}, \label{eq:min_spacing_constraint} && \\
		& \mathbf{r}_n, \mathbf{t}_m \in \mathcal{C}, \forall n, \forall m.\label{eq:aperture_constraint}
	\end{align}
\end{subequations}
}
Here, $w_{\mathrm{ISL}},w_{\mathrm{PSL}} \ge 0$ are the weighting coefficients for the \ac{isl} and \ac{psl}, respectively, such that $w_{\mathrm{ISL}} + w_{\mathrm{PSL}} = 1$. Constraint \eqref{eq:CommRate_constraint} guarantees that the downlink rate for each user remains above the threshold $\gamma_{\mathrm{min}}$, while \eqref{eq:Pill_constraint} enforces sufficient illumination throughout the ROI. Constraint \eqref{eq:tot_transmit_power} restricts the total transmit power to the available power budget $P_{\mathrm{total}}$. Finally, \eqref{eq:min_spacing_constraint} enforces the minimum inter-element spacing $d_0$ between all receive and transmit antennas, and \eqref{eq:aperture_constraint} confines all \acp{ma} to the feasible aperture $\mathcal{C}$.\par 

\subsection{Proposed Solution}
Problem $\mathrm{(P)}$ is challenging to solve due to (i) the coupling between antenna geometry and beamforming and the non-convexity of (ii) the \ac{psf}-based imaging objective, (iii) the communication-rate constraint \eqref{eq:CommRate_constraint}, and (iv) the minimum illumination power constraint \eqref{eq:Pill_constraint}. To tackle this problem, we propose a hybrid framework of \ac{ao} and metaheuristics that alternates between antenna-position optimization using \ac{pso} in subproblem $\mathrm{(P1)}$ and beamforming optimization using a \ac{ga} in subproblem $\mathrm{(P2)}$. Such approaches have proven effective for complex, high-dimensional optimization problems \cite{lahmeriUAVFormationResource2025a}. Further implementation details of the proposed algorithms are provided in the extended arXiv version \cite{luis}.
\subsection{Movable Antenna Placement Optimization} 
\label{subsec:MA_optim}
In subproblem $\mathrm{(P1)}$, the transmit and receive \ac{ma} positions, $\mathbf{T}$ and $\mathbf{R}$, respectively, are optimized for a fixed beamforming matrix \(\bm{\mathrm{W}}\). The resulting antenna placement optimization subproblem is given by 
\begin{equation}
\begin{aligned}
(\mathrm{P}1)\quad
\underset{\bm{\mathrm{T}},\bm{\mathrm{R}}}{\min}\quad
& J_{0}(\bm{\mathrm{T}},\bm{\mathrm{R}},\bm{\mathrm{W}}) \\
\mathrm{s.t.}\quad
& \eqref{eq:Pill_constraint},
\eqref{eq:min_spacing_constraint},
\eqref{eq:aperture_constraint}.
\end{aligned}
\label{eq:objective_SP1}
\end{equation}
Subproblem $\mathrm{(P1)}$ remains challenging to solve due to its complex, nonlinear, and highly non-convex \ac{psf}-based objective function $J_0$. Motivated by its success in solving similar antenna topology optimization problems \cite{jiangMASCOTMaskAssistedSpatial2025, lahmeriUAVFormationResource2025a}, we adopt \ac{pso} to solve $\mathrm{(P1)}$. 
In preliminary experiments, we observed that enforcing \eqref{eq:CommRate_constraint} during the geometry update can lead to only marginal position changes across iterations. A similar observation was reported in \cite{maMovableAntennaEnhanced2026}. Therefore, we omit this constraint in $\mathrm{(P1)}$ for placement optimization and enforce it in $\mathrm{(P2)}$ for the subsequent beamforming update, thereby ensuring the feasibility of the final solution. \ac{pso} employs $D_{\mathsf{PSO}}$ particles, each encoding a candidate topology via the \ac{2d} coordinates of all transmit and receive \acp{ma}, i.e., a vector in $\mathbb{R}^{2(M_{\mathrm{t}}+N_{\mathrm{r}})}$. Following \cite{lahmeriUAVFormationResource2025a}, a linearly decaying inertia weight is adopted, while particles crossing the aperture boundary are reflected back into $\mathcal{C}$. The illumination and minimum-spacing constraints are enforced using a parameter-free fitness function, following the constraint-handling principle in \cite{debEAcientConstraintHandling}.
\subsection{Beamforming Update by GA}

In subproblem $(\mathrm{P}2)$, the transmit beamforming matrix is optimized for a fixed antenna geometry
\begin{equation}
\begin{aligned}
(\mathrm{P}2)\quad
\underset{\bm{\mathrm{W}}}{\min}\quad
& J_{0}(\bm{\mathrm{T}},\bm{\mathrm{R}},\bm{\mathrm{W}}) \\
\mathrm{s.t.}\quad
& \eqref{eq:CommRate_constraint},
\eqref{eq:Pill_constraint},
\eqref{eq:tot_transmit_power}.
\end{aligned}
\label{eq:objective_SP2}
\end{equation}
Given the higher dimensionality and the non-convex objective and constraints, we solve \((\mathrm{P}2)\) using a continuous \ac{ga} \cite{hauptPracticalGeneticAlgorithms2004}.
Each chromosome encodes the magnitudes and phases of the beamforming coefficients, i.e., a vector $\bm{\mathrm{g}} \in [0,1]^{2\times M_{\mathrm{t}}(K+S)}.$ A population of \(D_{\mathsf{GA}}\) chromosomes is evolved using elitism, tournament selection, arithmetic crossover, and mutation. In contrast to the relaxed \ac{pso} update, all constraints in \((\mathrm{P}2)\) are strictly enforced using the feasibility-based constraint handling proposed in \cite{debEAcientConstraintHandling}. At each iteration, the best \(D_{\mathsf{kept}}\) chromosomes are retained, and only newly generated or mutated chromosomes are re-evaluated. 
\subsection{Overall AO Algorithm}
\label{subsec:alternating_scheme}
The proposed \ac{ao} algorithm combines the \ac{pso}-based \ac{ma} update and the \ac{ga}-based beamforming update within an outer iterative loop. Preliminary simulations showed that fully solving both subproblems using the stochastic optimizers \ac{pso} and \ac{ga} at every outer iteration incurs high computational cost and limits the exploration of the joint solution space due to the strong coupling between the antenna placement and beamforming problems. Therefore, we propose an adaptive \ac{ao} framework that dynamically controls the number of inner iterations to balance exploration and exploitation. In each \ac{ao} iteration $\alpha$, we maintain a \emph{current solution}
$\widetilde{\mathcal{X}} \triangleq
\{\widetilde{\mathbf{T}},\widetilde{\mathbf{R}},\widetilde{\mathbf{W}}\}$,
which initializes the subsequent updates, and a \emph{best solution}
$\mathcal{X}_{\star}$, which stores the best feasible solution found so far. Starting from $\widetilde{\mathcal{X}}$, the \ac{pso} optimizer generates a candidate antenna geometry $(\mathbf{T},\mathbf{R})$, after which the \ac{ga} optimizer refines the beamformer $\mathbf{W}$ to improve the sensing objective while satisfying the feasible set of problem $\mathrm{(P)}$, denoted by $\mathcal F$. To enhance exploration, up to $\eta$ consecutive \ac{ga} calls are allowed for beamforming refinement. The value of $\eta$ is adaptively determined using a counter $c$ that records the number of consecutive outer iterations for which \ac{pso} produces only a negligible geometry displacement relative to the current solution (see Line~\ref{alg_line:dgeo} of \textbf{Algorithm}~\ref{alg:alternating_optimization}). Thus, when a significant geometry change occurs, more \ac{ga} iterations are allocated to refine the beamformer and restore feasibility. Note that the {\em current solution} may accept slightly inferior feasible candidates within the acceptance margin $\Delta_{\mathsf{acc}}$ to enable controlled exploration, whereas the best solution is updated only upon improvement and is therefore monotonically non-increasing. The algorithm terminates after $\alpha_{\mathsf{AO}}$ outer iterations. These steps are summarized in \textbf{Algorithm}~\ref{alg:alternating_optimization},
with \(\operatorname{PSO}(\cdot)\) and \(\operatorname{GA}(\cdot)\) detailed in \cite{luis}. The worst-case computational complexity is 
\linebreaker{
\begin{equation}
\begin{aligned}
\mathcal{O}\Bigl(
&\alpha_{\mathsf{AO}}
\Bigl[
D_{\mathsf{PSO}}(T_{\mathsf{PSO}}+1)
+\eta_{\mathsf{max}}G D_{\mathsf{GA}}(T_{\mathsf{GA}}+1)
\Bigr] \\
&\times
\Bigl[
QM_{\mathrm{t}}^{2}N_{\mathrm{r}}^{2}
+Q^{2}M_{\mathrm{t}}N_{\mathrm{r}}
\Bigr]
\Bigr).
\end{aligned}
\end{equation}
}{
\begin{equation}
\mathcal{O}\Bigl(
\alpha_{\mathsf{AO}}
\Bigl[
D_{\mathsf{PSO}}(T_{\mathsf{PSO}}+1)
+\eta_{\mathsf{max}}D_{\mathsf{GA}}(T_{\mathsf{GA}}+1)
\Bigr] \times
\Bigl[
QM_{\mathrm{t}}^{2}N_{\mathrm{r}}^{2}
+Q^{2}M_{\mathrm{t}}N_{\mathrm{r}}
\Bigr]
\Bigr).
\end{equation}
}
where \(T_{\mathsf{PSO}}\) and $T_{\mathsf{GA}}$ denote the maximum numbers of \ac{pso} and \ac{ga} iterations, respectively.
\linebreaker{}{}
}
\linebreaker{
\versiontext{\section{Simulation Results}
We evaluate the proposed \ac{ma}-enabled \ac{isac} framework and the \ac{ao} algorithm using the simulation parameters summarized in Table \ref{tab:parameters}. The $K$ users are placed at the first $K$ positions in the ordered list
$\{[2,0,3]^\mathsf{T},\allowbreak
[-2,2,8]^\mathsf{T},\allowbreak
[-2,0.5,10]^\mathsf{T},\allowbreak
[-4,3,5]^\mathsf{T},\allowbreak
[4,-3,5]^\mathsf{T},\allowbreak
[-3,-4,8]^\mathsf{T},\allowbreak
[3,4,8]^\mathsf{T}\}\,\si{m}$. The spatial sampling interval is selected according to the expected cross-range resolution, \(\Delta \approx \frac{c_{0}R}{2f_{\mathrm{c}}D}=\SI{0.06}{m},\) such that the target pixel represents the desired mainlobe response, while all remaining pixels are included in the sidelobe evaluation.
We consider three fixed-array benchmarks: random antenna placement, a rectangular array, and a Mills Cross array \cite{zhugeStudyTwoDimensionalSparse2012, ahmedElectronicMicrowaveImaging}. For all fixed-array benchmarks, only the beamforming matrix is optimized using the same \ac{ga}-based beamforming procedure as in the proposed \ac{ao} framework. Furthermore, we consider a \ac{zf} benchmark that employs \ac{zf} precoding for communication, while optimizing only the sensing beamformer. Unless stated otherwise, all results are averaged over \(100\) independent realizations. Since no favorable antenna geometry or beamforming configuration is known a priori, the antenna positions for the proposed \ac{ao} algorithm and random-array benchmark are randomly initialized while satisfying \eqref{eq:min_spacing_constraint} and \eqref{eq:aperture_constraint}. For all schemes except \ac{zf}, the beamforming matrix is initialized as $\bm{\mathrm{W}}=\sqrt{P_{\mathrm{total}}/M_{\mathrm{t}}}\,\bm{\mathrm{I}}_{M_{\mathrm{t}}}$.\par
\versiontext{
\vspace{-0.3cm}
\begin{table}[h]
    \centering
    \caption{Simulation and algorithm parameters}
    \label{tab:parameters}
    \begin{tabular}{ll!{\vrule width 0.8pt}ll}
        \toprule
        \textbf{Parameter} & \textbf{Value} &
        \textbf{Parameter} & \textbf{Value} \\
        \midrule
         $M_{\mathrm{t}},N_{\mathrm{r}}$ & $8$ &
        $w_{\mathrm{ISL}}, w_{\mathrm{PSL}}$& $0.5$ \\
        $f_{\mathrm{c}}$ & $\SI{30}{GHz}$ &
        \ac{roi} size  & $\SI{1}{m^2}$ \\
        $z_{\mathrm{ROI}}$ & $\SI{3}{m}$ &
        $D$ & $\SI{0.25}{m}$ \\
        $Q$ & $289$ &
        $d_0$ & $\SI{5}{mm}$ \\
        $P_{\mathrm{total}}$ & $\SI{30}{dBm}$ &
        $P_{\mathsf{ROI}}$ & $\SI{-50}{dBm}$ \\
        $\sigma_{\mathrm{n}}^2$ & $\SI{-92.2}{dBm}$ &
        $\gamma_{\mathsf{min}}$ & $\SI{1}{\bit\per\second\per\hertz}$ \\
        $\alpha_{\mathsf{AO}}$ & $1000$ &
        $T_{\mathsf{PSO}}$ & $50$ \\
        $D_{\mathsf{PSO}}$ & $35$ &
        $T_{\mathsf{GA}}$ & $100$ \\
        $D_{\mathsf{GA}}$ & $32$ &
        $(\eta_{\min},\eta_{\max})$ & $(5,30)$ \\
        $c_1$ & $2$ &
        $c_2$ & $2$ \\
         $(\omega_{\min},\omega_{\max})$ & $(0.3,0.5)$ &
         $\mu$ & $0.05$\\
        $D_{\mathrm{kept}}$ & $16 $ &
        $\Delta_{\mathrm{acc}}$ & $0.001$ \\
        \bottomrule
    \end{tabular}
    \vspace{-0.2cm}
\end{table}

}{\linebreaker{\begin{table}[!t]
    \centering
    \caption{Simulation and algorithm parameters}
    \vspace{-0.2cm}
    \label{tab:parameters}
    \begin{tabular}{ll!{\vrule width 0.8pt}ll}
        \toprule
        \textbf{Parameter} & \textbf{Value} &
        \textbf{Parameter} & \textbf{Value} \\
        \midrule
        $M_{\mathrm{t}},N_{\mathrm{r}}$ & $8$ &
        $f_{\mathrm{c}}$ & $\SI{30}{GHz}$ \\
        \ac{roi} size & $\SI{1}{m^2}$ &
        $z_{\mathrm{ROI}}$ & $\SI{3}{m}$ \\
        $D$ & $\SI{0.25}{m}$ &
        $d_0$ & $\SI{5}{mm}$ \\
        $Q$ & $289$ &
        $P_{\mathrm{total}}$ & $\SI{30}{dBm}$ \\
        $P_{\mathsf{ROI}}$ & $\SI{-50}{dBm}$ &
        $\sigma_{\mathsf{Noise}}^2$ & $\SI{-92.2}{dBm}$ \\
        $\gamma_{\mathsf{min}}$ & $\SI{1.3}{\bit\per\second\per\hertz}$ &
        $\alpha_{\mathsf{AO}}$ & $1000$ \\
        $(T_{\mathsf{PSO}},D_{\mathsf{PSO}})$ & $(50,35)$ &
        $(T_{\mathsf{GA}},D_{\mathsf{GA}})$ & $(100,32)$ \\
        $(\eta_{\min},\eta_{\max})$ & $(5,30)$ & 
          $w_{\mathrm{ISL}}, w_{\mathrm{PSL}}$& $0.5$ \\
        \bottomrule
    \end{tabular}
    \vspace{-0.5cm}
\end{table}}{\begin{table}[ht]
    \centering
    \caption{Simulation and algorithm parameters}
    \vspace{-0.2cm}
    \label{tab:parameters}
    \begin{tabular}{ll!{\vrule width 0.8pt}ll}
        \toprule
        \textbf{Parameter} & \textbf{Value} &
        \textbf{Parameter} & \textbf{Value} \\
        \midrule
        $M_{\mathrm{t}},N_{\mathrm{r}}$ & $8$ &
        $f_{\mathrm{c}}$ & $\SI{30}{GHz}$ \\
        \ac{roi} size & $\SI{1}{m^2}$ &
        $z_{\mathrm{ROI}}$ & $\SI{3}{m}$ \\
        $D$ & $\SI{0.25}{m}$ &
        $d_0$ & $\SI{5}{mm}$ \\
        $Q$ & $289$ &
        $P_{\mathrm{total}}$ & $\SI{30}{dBm}$ \\
        $P_{\mathsf{ROI}}$ & $\SI{-50}{dBm}$ &
        $\sigma_{\mathsf{Noise}}^2$ & $\SI{-92.2}{dBm}$ \\
        $\gamma_{\mathsf{min}}$ & $\SI{1.3}{\bit\per\second\per\hertz}$ &
        $\alpha_{\mathsf{AO}}$ & $1000$ \\
        $(T_{\mathsf{PSO}},D_{\mathsf{PSO}})$ & $(50,35)$ &
        $(T_{\mathsf{GA}},D_{\mathsf{GA}})$ & $(100,32)$ \\
        $(\eta_{\min},\eta_{\max})$ & $(5,30)$ & 
          $w_{\mathrm{ISL}}, w_{\mathrm{PSL}}$& $0.5$ \\
        \bottomrule
    \end{tabular}
    \vspace{-0.5cm}
\end{table}}
}
\versiontext{}{\linebreaker{
\versiontext{
\vspace{-0.3cm}
\begin{figure}[ht]
\centering
\includegraphics[width=\figwidth]{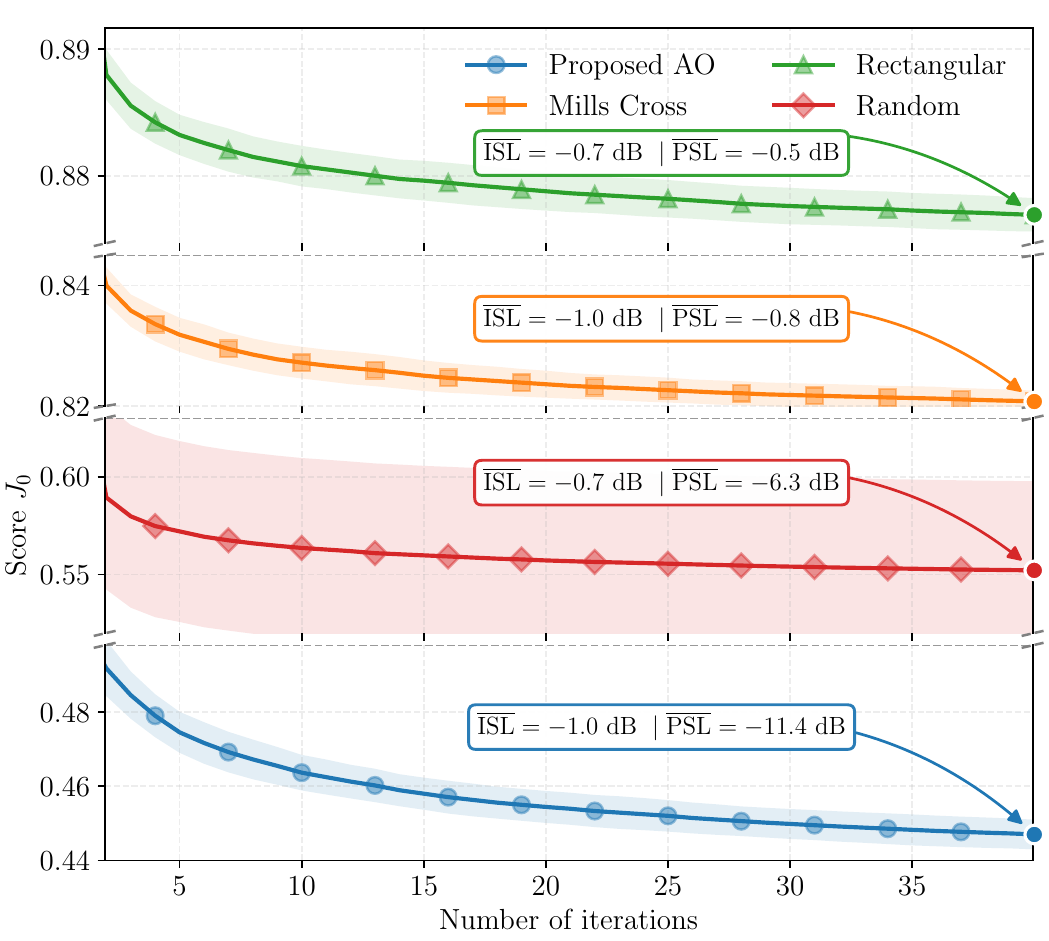}
\vspace{-0.3cm}
\caption{Convergence of the proposed \ac{ao} algorithm and the fixed-array benchmark schemes. Solid lines represent mean values, and shaded regions indicate one standard deviation.}
\vspace{-0.2cm}
\label{fig:convergence}
\end{figure}}{
\vspace{-0.4cm}
\begin{figure}[H]
\centering
\includegraphics[trim={0cm 0cm 0cm 0.2cm}, clip, width=\figwidth]{graphics/Convergence/convergence.pdf}
\vspace{-0.3cm}
\caption{Convergence of the proposed \ac{ao} algorithm and the fixed-array benchmark schemes. Solid lines represent mean values, and shaded regions indicate one standard deviation.}
\vspace{-0.3cm}
\label{fig:convergence}
\end{figure}}}{}}
Fig.~\ref{fig:convergence} compares the convergence behavior of the proposed \ac{ao} algorithm with the fixed-array benchmark schemes  for \(K=3\) users. The proposed \ac{ao} achieves the lowest objective value while maintaining low variability, with a substantial reduction during the first iterations, followed by gradual improvements. In contrast, the fixed-array benchmarks converge to considerably higher objective values, while the random-array benchmark achieves better performance but exhibits greater variability across independent runs. As each \ac{ao} iteration comprises both a \ac{pso}-based antenna-position update and a \ac{ga}-based beamforming update, the proposed algorithm entails a higher computational effort per iteration compared to the benchmarks.\par
\versiontext{
\versiontext{
\vspace{-0.3cm}
\begin{figure}[ht]
\centering
\includegraphics[width=\figwidth]{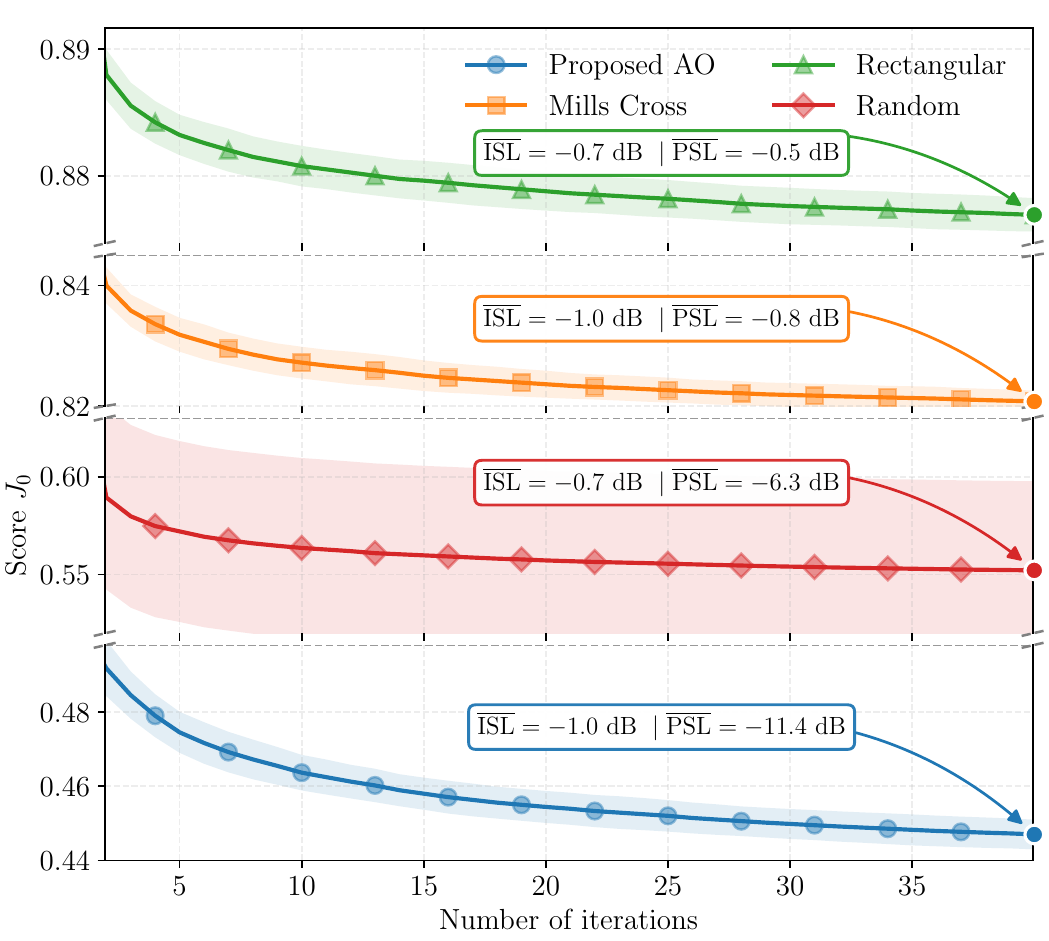}
\vspace{-0.3cm}
\caption{Convergence of the proposed \ac{ao} algorithm and the fixed-array benchmark schemes. Solid lines represent mean values, and shaded regions indicate one standard deviation.}
\vspace{-0.2cm}
\label{fig:convergence}
\end{figure}}{
\vspace{-0.4cm}
\begin{figure}[H]
\centering
\includegraphics[trim={0cm 0cm 0cm 0.2cm}, clip, width=\figwidth]{graphics/Convergence/convergence.pdf}
\vspace{-0.3cm}
\caption{Convergence of the proposed \ac{ao} algorithm and the fixed-array benchmark schemes. Solid lines represent mean values, and shaded regions indicate one standard deviation.}
\vspace{-0.3cm}
\label{fig:convergence}
\end{figure}}
\vspace{-0.2cm}
\begin{figure}[ht]
\centering
\includegraphics[trim={0cm 0.9cm 0cm 2cm}, clip, width=\figwidth]{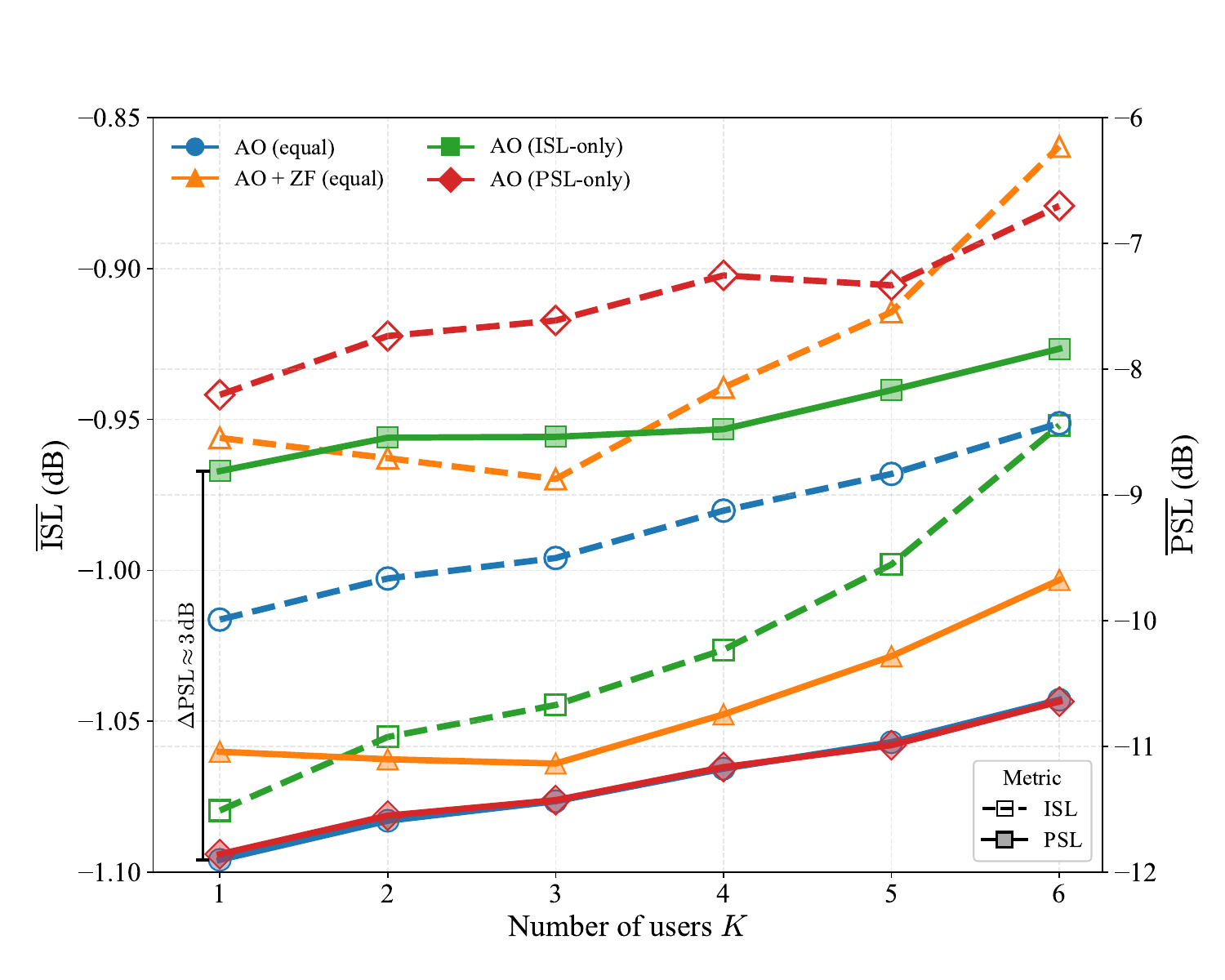}
\vspace{-0.3cm}
\caption{Average \ac{isl} and \ac{psl} versus the number of communication users $K$.}
\vspace{-0.3cm}
\label{fig:metrics_vs_users}
\end{figure}}{}
\versiontext{\begin{figure*}[!t]
\centering
\subfloat[Proposed]{
\includegraphics[trim={15.5cm 0cm 0cm 0cm} , clip, width=0.235\linewidth]{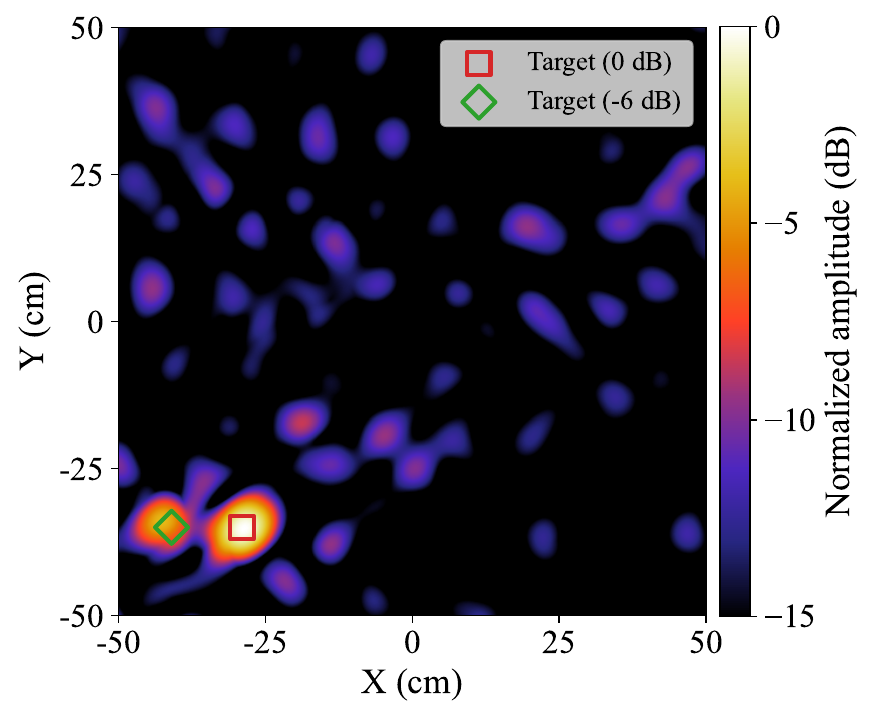}
\label{fig_first_case}}
\hfil
\subfloat[Random array]{
\includegraphics[trim={15.5cm 0cm 0cm 0cm} , clip, width=0.235\linewidth]{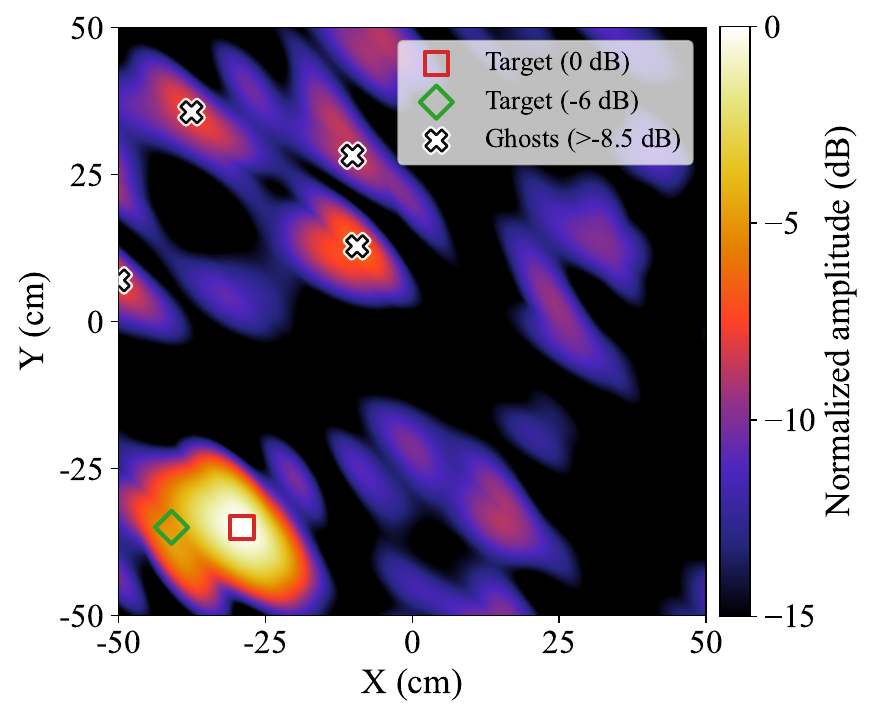}
\label{fig_second_case}}
\hfil
\subfloat[Mills Cross array]{
\includegraphics[trim={15.5cm 0cm 0cm 0cm} , clip, width=0.235\linewidth]{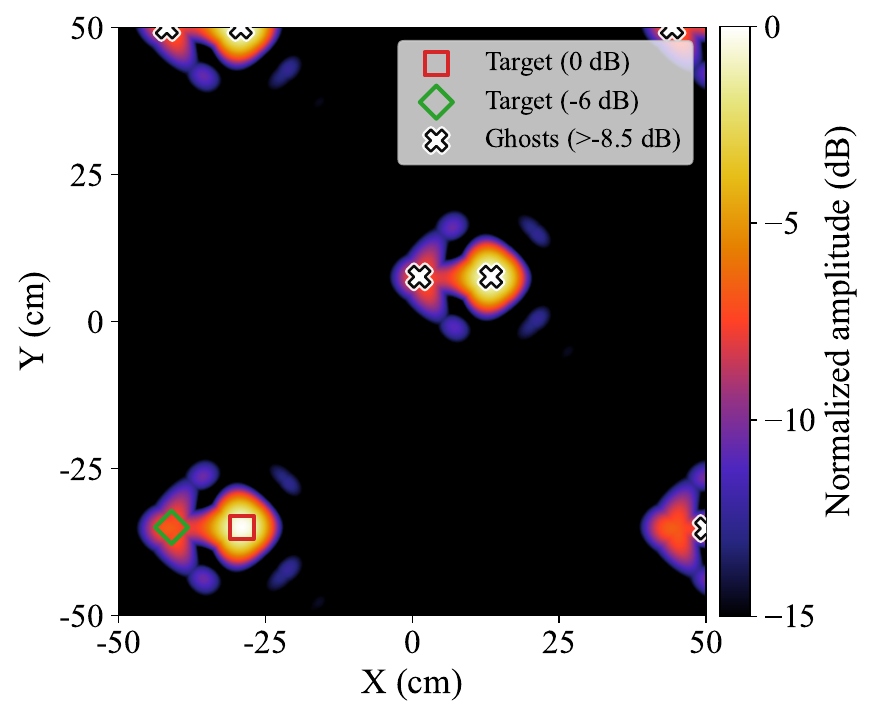}
\label{fig_third_case}}
\hfil
\subfloat[Rectangular array]{
\includegraphics[trim={15.5cm 0cm 0cm 0cm} , clip, width=0.235\linewidth]{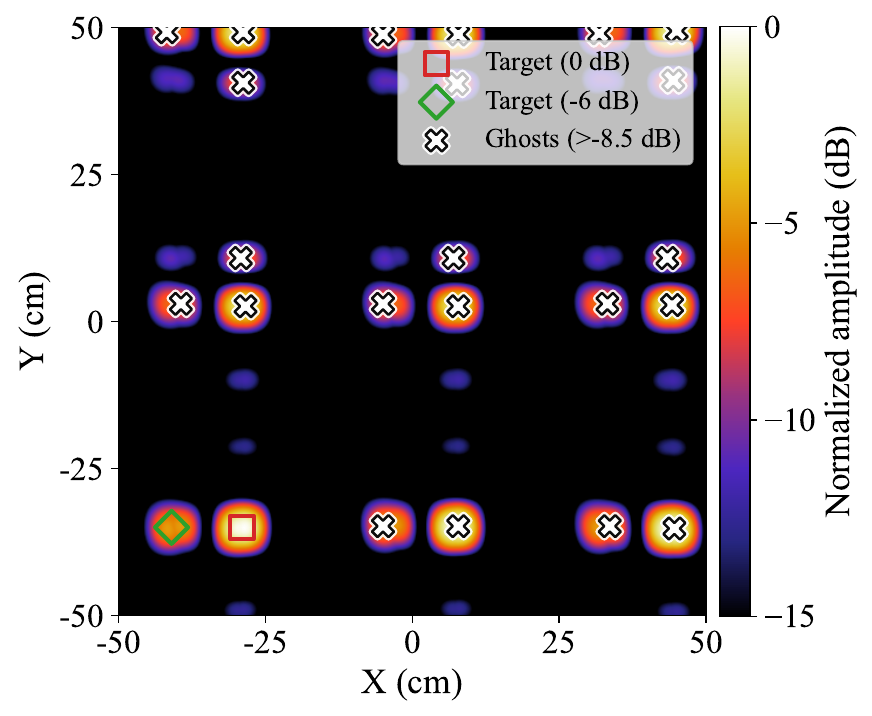}
\label{fig_fourth_case}}
\vspace{-0.1cm}
\caption{\ac{2d} imaging results for two closely spaced strong and weak point targets  located at $(\SI{-30}{cm}, \SI{-35}{cm}, \SI{3}{m})$ and $(\SI{-40}{cm}, \SI{-35}{cm}, \SI{3}{m})$, respectively.}
\vspace{-0.6cm}
\label{fig:image_x}
\end{figure*}}{}
Fig.~\ref{fig:metrics_vs_users} shows the \ac{isl} and \ac{psl} achieved by the proposed \ac{ao} for different objective weights and numbers of users $K$. Equal weighting achieves nearly identical \ac{psl} to \ac{psl}-only $(w_{\mathrm{PSL}} = 1)$ optimization, with minor differences due to stochasticity, while yielding a much lower \ac{isl}. The \ac{isl} advantage of \ac{isl}-only $(w_{\mathrm{ISL}} = 1)$ over equal weight optimization is most pronounced for small $K$, where more spatial \ac{dof} are available for imaging, but vanishes at $K=6$. However, this comes at the cost of an approximately $\SI{3}{dB}$ higher \ac{psl}, potentially causing stronger ghost targets and masking weak scatterers. Equal weighting thus provides a favorable compromise between total sidelobe energy and peak sidelobe suppression. Moreover, the increasing performance gap between the proposed \ac{ao} algorithm and the \ac{zf}-based scheme at larger $K$ highlights the growing benefit of joint beamforming and position optimization. Overall, both metrics deteriorate with increasing $K$, illustrating the trade-off between communication and imaging performance.\par
\versiontext{}{\linebreaker{\vspace{-0.2cm}
\begin{figure}[ht]
\centering
\includegraphics[trim={0cm 0.9cm 0cm 2cm}, clip, width=\figwidth]{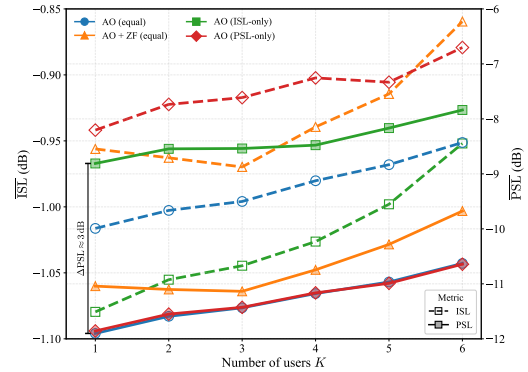}
\vspace{-0.3cm}
\caption{Average \ac{isl} and \ac{psl} versus the number of communication users $K$.}
\vspace{-0.3cm}
\label{fig:metrics_vs_users}
\end{figure}}{}}
Fig. \ref{fig:image_x} shows the imaging responses of two closely spaced scatterers with a $\SI{6}{dB}$ reflectivity difference near the edge of the \ac{roi}, and $K=3$ users. Compared with the Mills Cross and rectangular arrays, the random and proposed configurations avoid the pronounced grating lobes caused by regular and excessively large element spacings \cite{ahmedElectronicMicrowaveImaging}. However, the random array exhibits stronger sidelobe artifacts, creating ghost targets and overlapping target responses, making the two scatterers difficult to distinguish. In contrast, the proposed scheme produces two clearly distinguishable responses with reduced sidelobe levels, suppressing ghost-target artifacts and demonstrating the benefits of \ac{ma}-position optimization under communication constraints.

\section{Conclusion}
This paper demonstrated the potential of \acp{ma} for enhancing \ac{nf} \ac{isac} radar imaging through joint antenna-position and beamforming optimization. By directly minimizing the \ac{isl} and \ac{psl} of the \ac{2d} \ac{psf}, the proposed \ac{ao} framework achieves improved sidelobe suppression relative to the fixed-array geometries and \ac{zf}-based benchmarks. As a result, it reduces ghost targets and enables the distinction of closely spaced scatterers with different reflectivities while satisfying multi-user communication requirements. Our results further reveal that increasing the number of users degrades imaging performance, highlighting the trade-off between sensing and communication. These findings establish \acp{ma} as a promising additional degree of freedom for imaging-oriented \ac{isac} design. Future work will explore adaptive \ac{isl}/\ac{psl} weighting and more computationally efficient optimization strategies.}{
\begin{figure*}[!t]
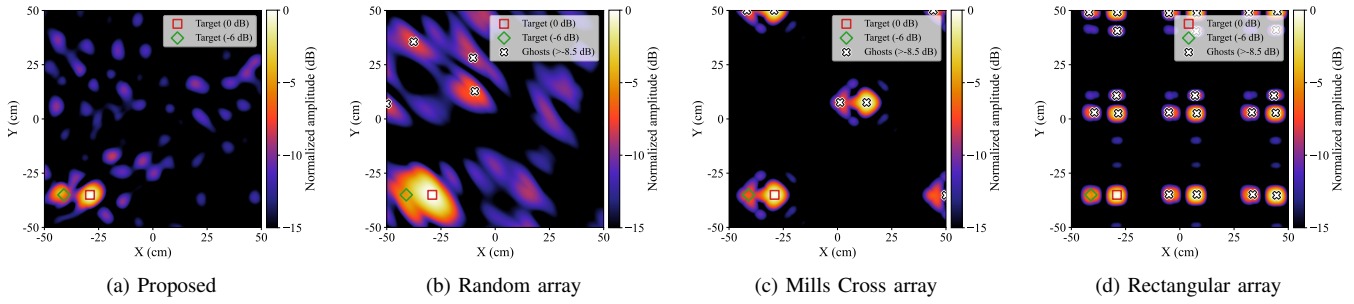

\centering
\subfloat[Proposed]{
\includegraphics[trim={15.5cm 0cm 0cm 0cm} , clip, width=0.235\linewidth]{graphics/Image_figures/AO/image.pdf}
\label{fig_first_case}}
\hfil
\subfloat[Random array]{
\includegraphics[trim={15.5cm 0cm 0cm 0cm} , clip, width=0.235\linewidth]{graphics/Image_figures/RANDOM/image.pdf}
\label{fig_second_case}}
\hfil
\subfloat[Mills Cross array]{
\includegraphics[trim={15.5cm 0cm 0cm 0cm} , clip, width=0.235\linewidth]{graphics/Image_figures/MC/image.pdf}
\label{fig_third_case}}
\hfil
\subfloat[Rectangular array]{
\includegraphics[trim={15.5cm 0cm 0cm 0cm} , clip, width=0.235\linewidth]{graphics/Image_figures/RECT/image.pdf}
\label{fig_fourth_case}}
\vspace{-0.1cm}
\caption{\ac{2d} imaging results for two closely spaced strong and weak point targets  located at $(\SI{-30}{cm}, \SI{-35}{cm}, \SI{3}{m})$ and $(\SI{-40}{cm}, \SI{-35}{cm}, \SI{3}{m})$, respectively.}
\vspace{-0.6cm}
\label{fig:image_x}
\end{figure*}

}
}{

\begin{figure*}[!t]
\centering
\subfloat[Proposed]{
\includegraphics[trim={15.5cm 0cm 0cm 0cm} , clip, width=0.235\linewidth]{graphics/Image_figures/AO/image.pdf}
\label{fig_first_case}}
\hfil
\subfloat[Random array]{
\includegraphics[trim={15.5cm 0cm 0cm 0cm} , clip, width=0.235\linewidth]{graphics/Image_figures/RANDOM/image.pdf}
\label{fig_second_case}}
\hfil
\subfloat[Mills Cross array]{
\includegraphics[trim={15.5cm 0cm 0cm 0cm} , clip, width=0.235\linewidth]{graphics/Image_figures/MC/image.pdf}
\label{fig_third_case}}
\hfil
\subfloat[Rectangular array]{
\includegraphics[trim={15.5cm 0cm 0cm 0cm} , clip, width=0.235\linewidth]{graphics/Image_figures/RECT/image.pdf}
\label{fig_fourth_case}}
\vspace{-0.1cm}
\caption{\ac{2d} imaging results for two closely spaced strong and weak point targets  located at $(\SI{-30}{cm}, \SI{-35}{cm}, \SI{3}{m})$ and $(\SI{-40}{cm}, \SI{-35}{cm}, \SI{3}{m})$, respectively.}
\vspace{-0.6cm}
\label{fig:image_x}
\end{figure*}
}

\renewcommand*{\bibfont}{\footnotesize}
\printbibliography
\end{document}